\documentclass[5p,times]{elsarticle}

\usepackage{lineno,hyperref}
\modulolinenumbers[5]

\journal{Future Generation Computer Systems}

\usepackage{amsmath,amssymb,amsfonts}
\usepackage{algorithm}
\usepackage{algorithmic}
\usepackage{textcomp}
\usepackage{lscape}

\begin{document}

\begin{frontmatter}

\title{Self-Assemble-Featured Internet of Things}
\tnotetext[mytitlenote]{DOI: \href{https://doi.org/10.1016/j.future.2020.05.012}{https://doi.org/10.1016/j.future.2020.05.012}.}
\tnotetext[mytitlenote]{\copyright  2020. This manuscript version is made available under the CC-BY-NC-ND 4.0 license (\href{http://creativecommons.org/licenses/by-nc-nd/4.0/}{http://creativecommons.org/licenses/by-nc-nd/4.0/}).}


\author[addressCNAM]{Fr\'ed\'eric Lemoine\corref{mycorrespondingauthor}}
\cortext[mycorrespondingauthor]{Corresponding author}
\ead{frederic.lemoine@cnam.fr}

\author[addressCNAM,addressTelecomParis]{Tatiana Aubonnet}

\author[addressTelecomParis]{No\"emie Simoni}

\address[addressCNAM]{Conservatoire National des Arts et M\'etiers (CNAM), CEDRIC, 292 rue Saint-Martin, Paris, France.}
\address[addressTelecomParis]{T\'el\'ecom-Paris, 19 Place Marguerite Perey, Palaiseau, France.}

\begin{abstract}
The Internet of Things supports various industrial applications. The cooperation and coordination of smart things are a promising strategy for satisfying requirements that are beyond the capacity of a single smart thing.
One of the major challenges for today's software engineering is the management of large and complex computing systems characterized by a high degree of physical distribution.
Examples of such systems arise in many application domains.
The number of connected devices grows from billions to hundreds of billions, so a maximum of automatisms must be integrated in IoT architectures in order to control and manage them. Software architects migrate to service oriented architecture and applications are now being constructed as service compositions.
Since each IoT device includes one or more microservices, the increasing number of devices around the user makes them difficult to assemble in order to achieve a common goal.
In this paper, we propose a self-assembling solution based on self-controlled service components taking into account non-functional requirements concerning the offered quality of services and the structuration of the resulting assembly. Its aim is to build and maintain an assembly of services (taking into account arrival of new peers or failure of existing ones) that, besides functional requirements, also fulfils global quality-of-service and structural requirements.
\end{abstract}


\end{frontmatter}

\section{Introduction}\label{sec_introduction}

The Internet of Things (IoT) vision defines a global network of interconnected services and smart objects that support human activities in everyday life using their sensing, computing and communication capabilities. IoT will enable global connectivity between devices, and people. The IoT links physical activities and real life with the virtual world \cite{khriyenko_user-assisted_2012}.

IoT will have a crucial role to play in the future, when human decision-making is necessary, especially in critical and urgent situations in which quality of service must be controlled. Some IoT-related critical systems are listed below:
\begin{itemize}
\item \textit{Energy and Utilities}: Increasing number of projects covering smart grid programmes, smart cities, and smart metering looking at ways to improve network efficiency and usability. Ensuring critical service requests (monitoring and control, power production, and water pressure monitoring) from trusted sources.
\item \textit{Automotive}: Sensors in vehicles provide even more data on things like environmental conditions, tyre pressure, engine performance and environmental conditions. These integrated, safe, and robust embedded systems will in a near future lead to self-driving cars. 
\item \textit{Railway (improvement of the management of rail networks)}: Widespread use of automation and sensor technologies, on tracks or trains, provides real-time data on resources in the network. This information can be used to predict when resources need repairing or replacing, reducing so the amount of time that rail resources are unavailable.
\item \textit{Healthcare}: Connected medical devices and applications are already creating an Internet of Medical Things which is contributing to better health monitoring and preventive care. From improving patients' experience through to connecting medical devices to physicians' smartphones, the IoT is transforming healthcare. Wearable devices and mobile medical applications that allow patients to measure and send their own biometric data, supplement patient treatment through remote monitoring and communication. The efficiency of healthcare facilities is also enhanced by improved tracking of medical devices, personnel and patients (Indoor Location Tracking). 
\item \textit{Aerospace (safety-critical applications)}: Sensors on aircraft create huge amounts of data for each flight, passenger management systems control huge amounts of complex personal data and air traffic control systems constantly monitor and manage plane flights, sharing data on a global scale. 
\end{itemize}

For any of these applications, failures might lead to serious injury and death (including on a large scale). As the number of objects, including sensors and actuators, increases, IoT becomes more and more complex and should be controlled, especially in these critical domains. Moreover, interactions between critical and non-critical systems in the IoT create problems for design engineers.  Consider an application to provide continuous tracking of a container of critical medical supplies (or transplant organs) from a first hospital to a second one far away. The container starts in a first hospital, then it is transferred to an ambulance, which travels by road to an airport. The container is then transferred into a plane, and, after landing, to a train, then to another ambulance, and finally arrives at a second hospital.  Many different IoT ecosystems were crossed, each complying with different standards and through a variety of communications links. 
Building this application in a way that patient lives are not put at risk, will be particularly challenging.

Applications are now being constructed as micro-service compositions \cite{art5:bluemix,art5:aws,art5:azure} integrating more and more functionalities.
Everyone can build a composition of services according to its needs. Software architects migrate to service-oriented architecture. We are in an age of services and micro-services are at the heart of architecture. If each IoT device includes one or more microservices, the increasing number of devices around the user makes them difficult to assemble in order to achieve a common goal.

The number of connected devices grows from billions to hundreds of billions, a maximum of automatisms must be integrated in the IoT architectures so as to control and manage them. 

A lot of IoT ecosystems rely on centralized, brokered communication models. All devices are identified, authenticated and connected through cloud servers that provide huge processing and storage capacities. Connection between devices will have to go exclusively through the internet, even if these devices are close to one another. These IoT ecosystems will not be able to manage a growing number of IoT devices. Cloud servers are a bottleneck that may disrupt the entire network.

A distributed service approach would solve the problem mentioned above, by spreading computational and storage requirements among billions of devices that will form the IoT networks of the future. 
Computing and storage are already widespread in many devices: from home to cars. Devices now carry as much computing power and connectivity as did the first smartphones. Connectivity and intelligence will be embedded in practically every object around us. Some services should stay close to the object instead of far in the cloud.
Many human-machine interactions will be replaced by machine-machine interactions.
IoT architectures will be more and more distributed and autonomous, acting in the best interest of the user, putting the user first, and designed for user-centric experiences. 

Furthermore, when a composition is defective from end to end, it is difficult to know which composite service is responsible. A fixed control centre, sufficiently powerful and capable of controlling the state of the entire system and manoeuvring its behaviour, is generally unreasonable. A maximum of automatisms should be placed on the objects themselves, thus unloading the cloud from this work. Service behavior should be controlled by the object itself rather than by the cloud. So we would gain efficiency and reaction time. 

The Internet of Things supports various industrial applications. The cooperation and coordination of smart things are a promising strategy for satisfying requirements that are beyond the capacity of any single smart thing.
One of the major challenges for today's software engineering is the management of large and complex computing systems characterized by a high degree of physical distribution.
Examples of such systems arise in many application domains, such as sensor networks, intelligent transportation systems, or ambient intelligence.

How can be helped the user and architect in building his application based on distributed services?

Since device failures and dynamics in the environment may imply the unavailability of components and devices at any time, adapting and maintaining such a composition is challenging.

How the QoS of such complex systems can be efficiently controlled?

Our motivation is to propose a self-assemble-featured architecture for IoT in a distributed service-oriented environment.
Its aim is to automatically build and maintain an assembly of services that besides functional requirements, fulfill global QoS and structural requirements. 
The concept of self-assembling covers multiple properties that can be regarded as the basic requirements for enabling the desired behaviors in such systems.

Analysis of existing solutions to meet the needs of IoT/cloud distributed systems, has led us to look at the evolution from component models to service models and their dynamic management. 
Component models provide a structured programming paradigm, and ensure a very good program re-usability. We focus here on generic and hierarchical component models because they make the design of large-scale systems easier.
On one side, QoS based approaches and adaptive contracts for services already exist (video services for example) \cite{cheng_adaptive_2015, wu_energy-efficient_2016} but they lack the compositional design and the management featured by components.
On the other side, dynamic management in time-critical IoT middleware is very compositional and expressive, but the support for an end to end QoS in the specific context of service-oriented components is weak. 
Time-critical IoT middlewares provide the QoS links but do not take into account the QoS nodes \cite{simpkin_constructing_2019, stefanic_switch_2019, ochian_overview_2014, cook_71_nodate}. 
To complete the modeling of the IoT ecosystem, we refer to the node/links model allowing us to have an architectural abstraction for dynamicity and a flexible engineering for a personalized user session.

Our proposal, presented in this paper, consists in using strongly structured components to provide a service oriented platform that eases the design and execution of applications with end to end QoS guarantees.
The dynamic management will be performed in a modular way. Each service (component) is responsible for its own adaptation and its own QoS while interacting with external services.

The extension of software services with autonomic capabilities has been suggested as a possible way to deal with this challenge.
One of the major challenges in IoT is the coordination that is necessary to align the behavior of different objects. 
Decentralization implies a style of coordination in which the objects cooperate as peers with respect to one another.
Self-assembly is an emergent property of decentralized systems.
It is the preferred mechanism for growth in the natural world, on scales ranging from molecular to macro-scales. It involves the assembling of components that are governed by a set of local interaction rules and lead to the formation of a global structure.
Understanding and applying this emergent property continues to be an important subject in the natural sciences, as well as in engineering and computer science.

The autonomic computing paradigm enables a system to automatically act in response to variations of operating conditions guaranteeing short reaction times and minimal human intervention. 
The approach we present in this paper is structured on a fully distributed implementation of several autonomic capabilities.
As an example, a possible scenario, to which we propose to answer, would be the following: A user buys an IoT device and brings it home. He connects it to his home network. The device is thus aware of its environment and detects the surrounding services and their features. These services are able to assemble themselves to achieve a common goal according to quality of service (QoS) requirements and thus respond to a user need (Control the home temperature, close the window, call emergency teams, etc.). If some IoT devices/services are appearing or disappearing, the system is able to reassemble itself in order to do its work and to continue to respect the QoS requirements.

To address the problem of complexity, collaboration, and QoS management of the growing number of IoT devices, we propose in this paper to use a distributed service architecture based on our service composition entity. We extend it with autonomic capabilities, and we introduce a way to automatically assemble a set of IoT device microservices in order to achieve a common goal. We will show that our solution is able to dynamically perform the composition and thus react to changes in its execution context.
We propose a self-assembling solution based on self-controlled service components taking into account non-functional requirements concerning the offered QoS and the structuration of the resulting assembly. Its aim is to build and maintain an assembly of services (taking into account arrival of new peers or failure of existing ones) that, besides functional requirements, also fulfills global quality-of-service and structural requirements. It is able to drive the system and manage the selection, among the set of functionally feasible assemblies, of an assembly that fulfills these requirements.

After enumerating in Section \ref{sec_properties}, the properties of Self-Assemble-Featured IoT, presenting our service composition entity on which we base our approach (Section \ref{sec_previous_works}) and reviewing related works (Section \ref{sec_related_works}), we present our self-assembly algorithm (Section \ref{sec_proposals}).
Section \ref{sec_case_studies} presents a case study. Finally, a conclusion (Section \ref{sec_conclusion}) ends the article.

\section{Properties of Self-Assemble-Featured IoT}\label{sec_properties}

The concept of self-assembling covers multiple properties that can be regarded as the basic requirements for enabling the desired behaviors in such systems. A qualified implementation of a self-assembling scheme should meet the following requirements that will be part of the future Internet of Things: (i) distributed infrastructure, (ii) autonomous, and (iii) efficient collaboration.

\begin{enumerate}
\item \textbf{Distributed infrastructure.} For the IoT, which is characterized by highly evolved heterogeneity, a fixed controlling center capable and powerful enough to control the status of the whole system and manoeuvre its behaviors is generally unreasonable. So, a self-assembling scheme based on the distributed infrastructure may be much more suitable for the IoT. 
\item \textbf{Autonomous.} Each node in the IoT has to be implemented with a certain degree of autonomy. Such nodes can therefore collect information about their respective environments and make decisions for their own behaviors. In this case, the data (including instructions and information) transmitted amongst nodes in a fully distributed form, is restricted into a limited area, which decreases heterogeneity of the whole network. Such distributed architecture also enables the self-assembling scheme to be scalable. 

Our approach is organized with respect to several autonomic capabilities \cite{art5:Lalanda:2013:ACP:2500995}. 
We recall the definitions for our approach in Table \ref{table_autonomic_properties}.
\begin{table*}[h]
\centering
\caption{Autonomic properties covered by our approach.}
\label{table_autonomic_properties}
\begin{tabular}{clc}
\hline
Properties & Definitions & See Section\\
\hline Self-defining & \begin{tabular}{l}                                               A system's ability to describe itself to other\\
                                                                                        systems. An autonomic system may need to\\
                                                                                        understand and interpret other systems'\\
                                                                                        descriptions.\end{tabular} & \ref{sec_dnssd}\\
\hline \begin{tabular}{l}Self-control\\Self-diagnosis\end{tabular} & \begin{tabular}{l} A system's ability to analyse itself in order\\
                                                                                        to identify existing problems or to anticipate\\
                                                                                        potential issues.\end{tabular} & \ref{sec_previous_works}\\
\hline Self-monitoring & \begin{tabular}{l}                                             A system's ability to retrieve information on its\\
                                                                                        internal state and behaviour, whether globally,\\
                                                                                        or for any of its constituent elements.\end{tabular} & \ref{sec_previous_works}\\
\hline Self-assembled & \begin{tabular}{l}                                              A system's property of being automatically formed\\
                                                                                        via the distributed assembly of multiple\\
                                                                                        independent elements, which become the system's\\
                                                                                        constituent elements.\end{tabular} & \ref{sec_algorithmic_model}\\
\hline Self-simulation & \begin{tabular}{l}                                             A system's ability to test and evaluate scenarios\\
                                                                                        without affecting the executing system (e.g. it\\
                                                                                        should not impact provided services). This allows\\
                                                                                        replying to 'what would happen if' questions and\\
                                                                                        hence facilitate the selection of self-adjusting actions.\end{tabular} & \ref{sec_benefits}\\
\hline Self-adjusting & \begin{tabular}{l}                                              A system's ability to modify itself during runtime,\\
                                                                                        including modifications to its internal structure.\end{tabular} & \ref{sec_benefits}\\
\hline Self-adapting & \begin{tabular}{l}                                               A system's ability to modify itself (self-adjust)\\
                                                                                        in reaction to changes in its execution context or\\
                                                                                        external environment, in order to continue to meet\\
                                                                                        its business objectives despite such changes.\end{tabular} & \ref{sec_benefits}\\
\hline \begin{tabular}{l}Self-healing\\Self-repair\end{tabular} & \begin{tabular}{l}    A system's ability to recover from the failure of\\
                                                                                        any of its constituent elements (service) or to\\
                                                                                        predict and prevent the occurrence of such failures.\end{tabular} & \ref{sec_benefits}\\
\hline
\end{tabular}
\end{table*}
Services have to be aware of their environment and therefore of their surrounding services. They need intelligent service discovery process. Services have to broadcast their features (function, offered/nominal QoS, threshold value from which the business component stops responding) to the others (\textbf{self-defining}). A method to implement this service discovery process is proposed in Section \ref{sec_dnssd}.
We recall, in Section \ref{sec_previous_works}, that our component fulfills \textbf{self-control/self-diagnosis, and self-monitoring} requirements.
In Section \ref{sec_algorithmic_model}, we present a self-assembling algorithm allowing self-controlled services to be assembled to compose an application (\textbf{self-assembled}). This algorithm test and evaluate scenarios to choose the more suitable for using taking account QoS constraints (\textbf{self-simulation}).
We show in Section \ref{sec_benefits} that our self-assembling algorithm, with its abilities to reassemble components at any time, supports the \textbf{self-adjusting, self-adapting, and self-healing (self-repair)} benefits.

\item \textbf{Efficient collaboration} based on ubiquitous data exchanging and sharing. We suppose in this paper that our components can exchange and share data in an efficient way. This point is out of the scope of this paper. Several approaches have been proposed in literature like "Blockchain" which brings security. A blockchain \cite{art5:samaniego_internet_2017,art5:zheng_overview_2017}, or chain of blocks, is the implementation of a data storage and transmission technology without a control organ. Technically, it is a distributed database whose information, sent by users, is checked and grouped in blocks, linked and secured using cryptography. Our algorithm requires that each node maintains and disseminates global state information consisting of the whole assembly of services and computed link QoS. The amount of data to be shared is small. This can be done with use of blockchain technology but other simplest solutions may be sufficient.  
\end{enumerate}

The following section presents our service composition entity on which we base our approach.

\section{Previous Works}\label{sec_previous_works}

At the service era, the service is the center of architecture, to profit by all the benefits expected from this concept, we have proposed in \cite{art5:aubobost} a component called "Self-Controlled service Component (SCC)", the description of which we recall. 
A component encapsulates a micro-service and only one. These services can be of different sizes and natures: real-time image analysis service (face recognition, character recognition, barcode recognition) in the case of the Internet of Things, algorithms (sorting, encryption), image capture, etc.
To describe the behavior of our components and allow an homogeneous quality of service management, a generic QoS model \cite{art5:DBLP:AubonnetS13} is defined. Four criteria are proposed to describe the QoS: availability, reliability, time, and capacity.
\begin{itemize}
\item \textbf{Availability} represents the accessibility rate of the service component.
\item A system is said to be \textbf{reliable} when its behavior, over a given duration, is in conformity with that expected.
\item \textbf{Time} represents the time required for request processing.
\item \textbf{Capacity} represents the maximum load that the service component can handle.
\end{itemize}
This proved to be useful and sufficient in all the practical cases we studied.
To increase the structural decomposition and the reuse of non-functional QoS components, its internal functions have been separated and an architecture that separates the monitoring and QoS functions of the remaining functions called "control" has been proposed.
This model has been specified in the OpenCloudware project \cite{art5:OpenCloudware} to address the behavioral aspects through QoS.
\begin{figure}[!t]
\centering
\includegraphics[width=0.48\textwidth]{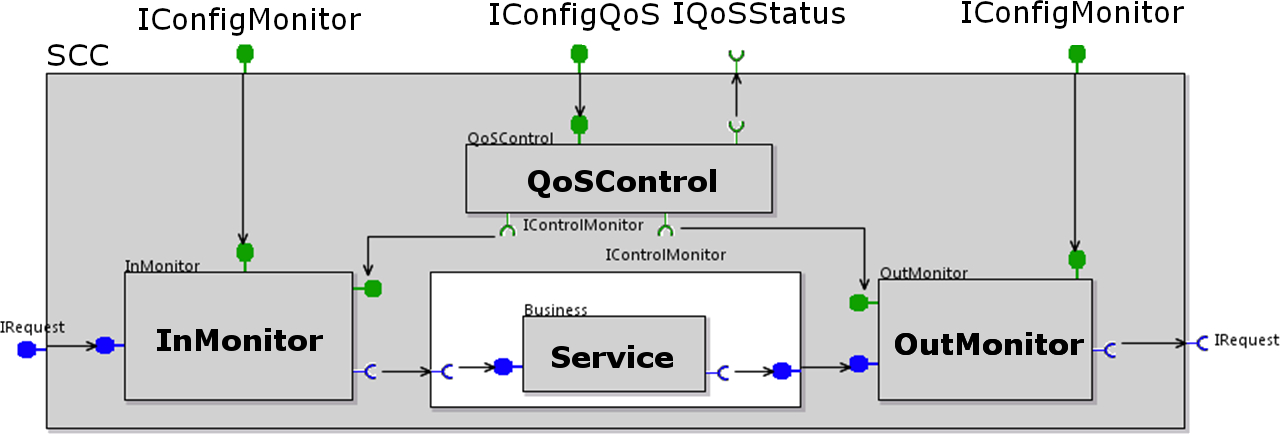}
\caption{Self-Controlled service Component (SCC)}
\label{fig_SCC_light}
\end{figure}

The membrane of our SCC includes (Figure \ref{fig_SCC_light}):
\begin{itemize}
\item Input monitoring (InMonitor) and output monitoring (OutMonitor) components. They play an interceptor role. Incoming service requests are intercepted and transmitted (unchanged) to the functional component via the corresponding internal interfaces. The OutMonitor intercepts outgoing service requests. They provide measurement information on the flow that they intercept.
\item A QoS component (QoSControl), associated with the business component.
\item A non-functional interface (client) for QoS control (IQoSStatus), by which it will send the information in case of violation of QoS contracts, i.e. "InContract" notifications when the behaviour is in conformity with the contract or "OutContract" otherwise.
\item A non-functional interface (server) of configuration (IConfigQoS, IConfigMonitor), whose role is to receive component configuration commands.
\end{itemize}
The QoSControl component checks the current behaviour of the resource and its conformity with the contract. For this, it regularly requests to the monitors (InMonitor and OutMonitor) the parameter values. It compares each current value to the corresponding threshold value not to be exceeded. It sends an OutContract notification if the current value is less (or more) than the threshold value. In this case the dynamic management consists in replacing immediately the failing component by a ubiquitous service fulfilling the requirements. Otherwise, it sends an InContract notification. We obtain an SCC component, self-monitored and self-controlled. The sub-components of the membrane (monitors and QoS) are activated in order to perform monitoring of the quality of service and to notify its degradation. 
This component has been designed so that it can be composed with others in order to form a self-controlled composition. An application is a service composition. The service composition includes different SCC called during the user session.

The SCC component has proved its value in the Cloud \cite{art5:OpenCloudware,art5:DBLP:AubonnetS13,art5:aubonnet:hal-01180627} and in the IoT \cite{lemoine_iot_2020}. The SCC architecture has many advantages: Our SCC component allows self-control inside by signaling dysfunction (out of contract) and automatic reacting outside. Since we are as close as possible to the functional component, the analysis is faster, more relevant, and reaction times are minimized. The analysis is done on the site itself. Only its result is sent so that the volume of data exchanged and thus the communication resources needed are extremely small. The code is simplified and hence requires less computing resources. Monitoring and controlling components are generic so they are independent of the functional component and may be present at all levels of architecture. They are not intrusive because they are external to the functional component. They are inside the service component membrane and operate in parallel with the functional component. They have no effects on the second. The QoS of each component (hardware or software) is measured, allowing better diagnostic of various dysfunctions whereas most existing tools monitor network traffic or central processing unit usage when they should monitor the functional component performance. At each addition/removal of a functional component, a monitoring and controlling component is therefore added/removed (\textit{Scalability, Elasticity}).

The service is defined by an offered/nominal QoS and a threshold value. These values are used by our algorithm (Section \ref{sec_algorithmic_model}). Offered/nominal QoS is the QoS (processing time for example) provided by the service under normal operating conditions. It is a value given by the service provider in a service level agreement or in its provided services catalogue. It is computed under resource conditions of the underlying level. The threshold value is the number of simultaneous requests not to exceed in order to comply with the nominal preceding defined QoS. These values are computed with a calibration procedure \cite{art5:lemoine:MaaS}.

The following section surveys the related works.

\section{Related Works}\label{sec_related_works}

In literature, the term "self-assembling" mainly concerns robotic, mechatronic, chemistry and materials science domains. There are few articles in the field of service-oriented architecture. Assembly is often done manually by a software architect or sometimes computed automatically. 

Methodology for middleware and technology selection is detailed as follows. We surveyed 50 middlewares and technologies and made an initial selection of 25 of the most advanced approaches that deal with service discovery and assembly with a preference for the self-assembly property.
Assembly can be done in different ways: (i) manually with a visual editor for example, (ii) automatically with an algorithm or predefined scripts (only once in general at the design phase or at demand), and (iii) dynamically (at any time with respect to context changes).
Assembly can be done in a centralized (by a central entity) or decentralized way (by the devices themselves).
We have chosen to take into account the control of QoS too (QoS aware). 
The choice has been restricted to a selection of 16 representative solutions: Tile-based approach, MACODO, Schuhmann et al., FlashMob, MOSDEN, UbiROAD, Calvin, Paraimpu, SENSEI, Node-RED, CHOReOS, SenseWrap, SOCRADES, Ubiware, extended GSN, KASOM (Table \ref{table_related_works}).

Article \cite{art5:cavallaro_tile-based_2010} presents an approach allowing to develop a service-oriented system, based on a model called service tiles, by building an assembly of service components that accomplishes a given goal. The assembly is computed automatically: with the specification of a subset of the whole system, a few constraints, and the goals that the application should fulfill.

MACODO \cite{art5:weyns_macodo_2010_middleware,art5:weyns_macodo_2010_organization} uses a partially distributed architecture based on a master-slave scheme. The master has a complete knowledge of the assembly state and controls the dynamics in a centralized way. Masters of different assemblies can cooperate to achieve a given goal.

Article \cite{art5:schuhmann_adaptive_2013} presents algorithms for homogeneous and heterogeneous environments whose goal is to choose the most efficient assembly method for a given environment while minimizing the assembly time. The organization latency is reduced by caching and reusing partial application assemblies.

FlashMob \cite{art5:sykes_flashmob:_2011} is based on dynamic service assembly and requires a backtracking phase to explore alternative solutions in case the assembly fails and has no global QoS goal. Its self-assembly procedure is decentralized. Global state information dealing with the whole assembly is disseminated among the services.

MOSDEN \cite{perera_mosden_2014} supports a sensing as a service model \cite{perera_sensing_2014}. It is an IoT middleware for resource constrained computational mobile devices. MOSDEN can collect data from multiple different sensors and process them together. It is fully compatible with Global Sensor Network Middleware that runs on the cloud. Sensor discovery and service composition are not automated.

UbiROAD \cite{terziyan_ubiroad_2010} is a specialized platform for smart traffic environments. Semantic technologies are the basis for the discovery of heterogeneous resources and data integration across multiple domains. They are used both for descriptive specification of services delivered by the resources and for prescriptive specification of the expected behavior of resources as well as of the integrated system. UbiROAD guarantees a high level of safety while offering personalization, dynamic behaviour, customization, and autonomy of services and ensures context-awareness, and adaptive/reconfigurable composition. 

Calvin \cite{calvin,persson_calvin_2015} is an open-source IoT middleware from Ericsson. IoT applications building is based on actors which are reusable software components that may represent a device, a computation, or a service. It comprises both a development framework for IoT application developers, and a runtime environment that handles the running application. Compositions are made by writing scripts called CalvinScript. Basically, an application consists of actor instances and connections between the ports of the actors, forming a data flow graph. To allow the reuse of scripts, it is also possible to define components. An application script can also contain deployment rules.

Paraimpu \cite{paraimpu} is a social aware IoT middleware that allows consumers to add, use, share, and interconnect RESTful IoT services whether physical or virtual. Things are mapped to abstract concepts of sensors or actuators. Providing connection abstraction between things, allows users to compose simple IoT applications via Javascript.

The SENSEI \cite{Tsiatsis2010} middleware includes context services, context model, actuation tasks, and dynamic service composition of both primitive and advanced services. 

Node-RED \cite{nodered} is an open-source IoT middleware platform from IBM. Its key advantage is a visual tool that simplifies the composition of IoT devices especially if the node for the IoT device is already developed and published by others.

CHOReOS \cite{autili_choreos_2014} composes distributed services by considering a global specification, called Choreography, of the interactions between the participant services. It enables large scale compositions or choreographies of QoS-aware, and heterogeneous services in IoT. It includes i) extensible service discovery to manage protocols and processes for discovery of services and things, and ii) executable service composition to coordinate the composition of services and things. Semantic thing-based service compositions are automatically executed, with no involvement from endusers. 

The SenseWrap \cite{evensen_sensewrap_2009} middleware combines the Zeroconf \cite{zeroconf} protocols with hardware abstraction using virtual sensors. A virtual sensor provides transparent discovery of resources, mainly sensors, through the use of Zeroconf protocol. Applications can thus use it in order to discover sensor-hosted services.  It also provides a standardized communication interface to hide the sensor-specific details. This interface depends on sensor modeling and custom wrappers. Virtualization is applied only to sensors, not to actuators or computing resources which makes it unsuitable for ultra-large scale IoT environments and heterogeneous networks.

The SOCRADES \cite{socredes_2010} middleware architecture consists of a layer for application services and a layer for device services (service discovery, monitor, device and service lifecycle management). By using devices profile for web services (DPWS), it abstracts physical things as services. Its cross-layer service catalogue supports service composition, but it may not be fully dynamic, since composition relies on predefined building blocks.

Ubiware \cite{palade_middleware_2017} is able to create flexible, autonomous, and complex industrial systems. It supports composition, invocations, monitoring, automatic resource discovery, and execution.

Global Sensor Networks (GSN) \cite{gsn_2014} is a service-based IoT middleware that proposes virtual sensor abstraction and aims to provide a uniform platform for integration and deployment of heterogeneous IoT devices. Users and developers specify XML-based deployment descriptors to deploy a sensor. GSN does not support composition of multivendor devices via the XML descriptor.  The extended GSN \cite{calbimonte_xgsn_2014} provides a limited composition capability.

KASOM \cite{corredor_knowledge_aware_2012} is a knowledge-aware and service-oriented middleware. Its aim is to offer advanced and enriched pervasive services to every user connected to the Internet. KASOM implements mechanisms and protocols that allow managing the knowledge generated in pervasive embedded networks in order to expose it to Internet users in a readable way.  Its architecture includes communication services (resource monitor), and knowledge management services (context resources and service composition rules). This middleware offers discovery, registration, orchestration, and composition of services. It has proven its qualities in terms of reliability, efficiency, and response time in the field of health management. However, because of predefined service composition rules provided by in-network agents, it does not provide dynamic service composition in IoT infrastructure. 

An application (service composition) is usually monolithic. 
It offers features created to meet a set of needs even if the user does not use them all.
A dynamic assembly would create a customized application according to the user's current need. 
It would integrate a service only if necessary and would thus gain in performance.
The increasing number of IoT devices, however, implies the problem of their management to achieve specific common objectives.
A fixed controlling center capable and powerful enough to control the status of the whole system and manoeuvre its behaviors is in general unreasonable. The assembly has to be decentralized.
The question is then how to coordinate local services among IoT devices in order to accomplish a goal with global QoS evaluation?

Table \ref{table_related_works} summarizes the analysis and categorization of these works.

None of the afore mentioned approaches can be simultaneously dynamic, decentralized, covering non-functional and structural constraints and integrates self-control mechanisms as defined in this paper.

\begin{table*}[h]
\centering
\small
\caption{Analysis of works related to assembly of services.}
\label{table_related_works}
\begin{tabular}{@{}lllll}
\hline\noalign{\smallskip}
 & Assembly & Service discovery & QoS control & Decentralized / centralized\\
 \noalign{\smallskip}\hline\noalign{\smallskip}
Tile-based approach \cite{art5:cavallaro_tile-based_2010} & Automatic & NI & No & NI\\
MACODO \cite{art5:weyns_macodo_2010_middleware,art5:weyns_macodo_2010_organization} & Dynamic & NI & No &\begin{tabular}{@{}l}Partially distributed. A\\master controls the\\ dynamics in a centralized\\way.\end{tabular}\\
Schuhmann et al. \cite{art5:schuhmann_adaptive_2013} & Dynamic & NI & No & NI\\
FlashMob \cite{art5:sykes_flashmob:_2011} & Dynamic & NI & No global QoS goal. & Decentralized\\
MOSDEN \cite{perera_mosden_2014} & Manual. Not automated & Not automated & NI & Cloud centric\\
UbiROAD \cite{terziyan_ubiroad_2010} & Adaptive/reconfigurable & \begin{tabular}{@{}l}Yes with descriptive\\specifications.\end{tabular}  & \begin{tabular}{l}Prescriptive specification of\\the expected behaviour.\end{tabular} & NI\\
Calvin \cite{calvin,persson_calvin_2015} & Made by writing scripts & NI & No & NI\\
Paraimpu \cite{paraimpu} & Manual. Via Javascript & NI & No & Cloud centric\\
SENSEI \cite{Tsiatsis2010} & Dynamic & NI & No & NI\\
Node-RED \cite{nodered} & Manual. Via a visual tool & NI & No & Cloud centric\\
CHOReOS \cite{autili_choreos_2014} & Computed automatically & \begin{tabular}{@{}l}Discovery of services\\and things.\end{tabular} & Yes & Decentralized\\
SenseWrap \cite{evensen_sensewrap_2009} & Manual. Not automated & Yes & No & Decentralized\\
SOCRADES \cite{socredes_2010} & \begin{tabular}{@{}l}Not fully dynamic,\\predefined building blocks\end{tabular} & Yes & No & NI\\
Ubiware \cite{palade_middleware_2017} & Not fully dynamic & Yes & No & Decentralized\\
extended GSN \cite{calbimonte_xgsn_2014} & \begin{tabular}{@{}l}Limited composition\\capability\end{tabular} & NI & No & NI\\
KASOM \cite{corredor_knowledge_aware_2012} & \begin{tabular}{@{}l}Not dynamic, predefined\\service composition rules\end{tabular} & Yes & \begin{tabular}{@{}l}Reliability, efficiency,\\and response time.\end{tabular} & NI\\
\noalign{\smallskip}\hline
\end{tabular}
\\No information (NI)
\end{table*}

In the following section, we present our self-assembly algorithm.

\section{Self-assembly Algorithm}\label{sec_proposals}

In Section \ref{sec_dnssd}, we describe a method for a service to describe its capabilities especially his QoS nominal/offered to others in order to achieve a requested service composition.
In Section \ref{sec_link_qos}, we show that, using our component, we are able to compute an estimation of the link QoS. 
In Section \ref{sec_algorithmic_model}, we define our model, introduce the terminology and notation used in the rest of the paper, present our assembling core algorithm, and enumerate the benefits of our algorithm.
Finally, the limitations of the algorithm on a real IoT platform are estimated (Section \ref{sec_resource_consumption}). 

\subsection{Service Self-Definition and Surrounding Awareness}\label{sec_dnssd}

In this section, a method, for a service to describe its capabilities, is proposed. Each service has to broadcast its features (function, offered/nominal QoS, threshold value) to the others. This is proposed as a proof of concept. Our proposition is based on Network Service Discovery (NSD) protocol which we propose to extend. NSD gives to any application access to services that other devices provide on a local network. 
Devices that support it include webcams, printers, HTTPS servers, and mobile devices. Provided services include web server, file sharing, file transfer, printing, real time streaming, sound server, etc. 
It's based on the Domain Name System-based Service Discovery (DNS-SD) mechanism \cite{art5:rfc6763}, which allows the user's app to request services by specifying a type of service and the name of a device instance that provides the desired type of service. DNS-SD is supported both on Android and on other mobile platforms.
\begin{figure}[!t]
\centering
\includegraphics[width=0.42\textwidth]{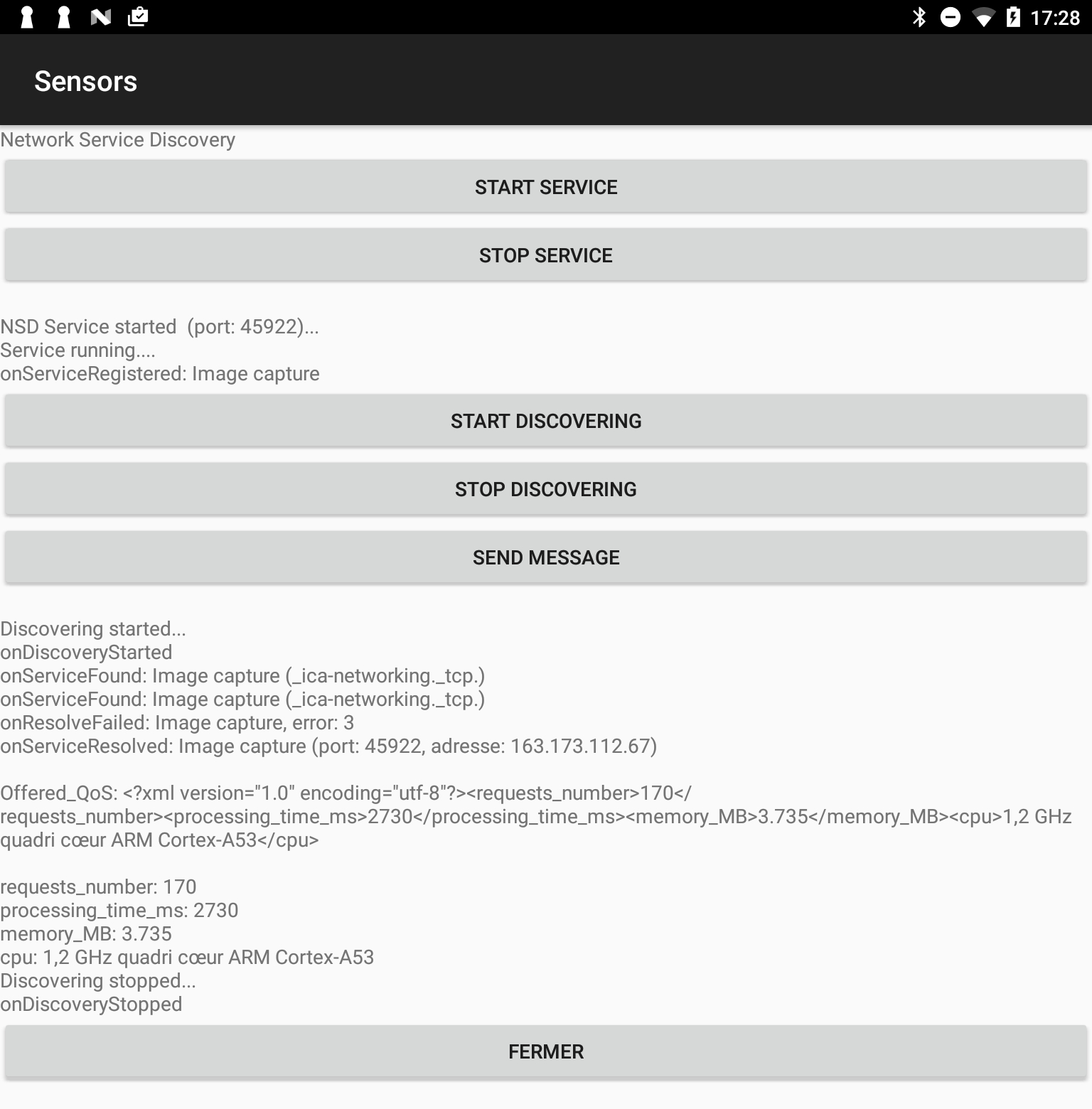}
\caption{Self-definition of SCC using NSD on Android}
\label{fig_DNS_SD}
\end{figure}
\begin{figure*}[!t]
\centering
\includegraphics[width=1.0\textwidth]{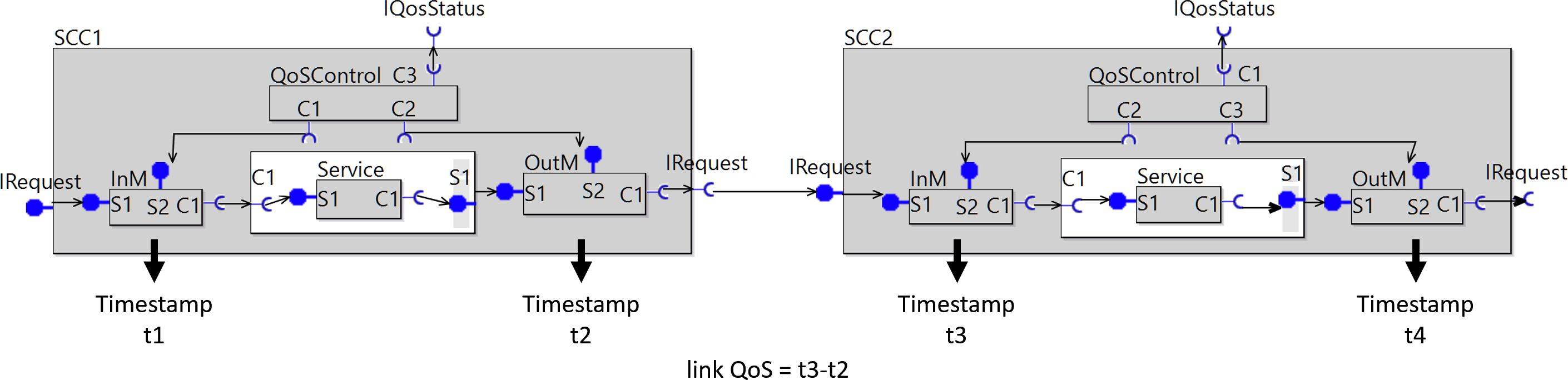}
\caption{Estimation of the link QoS.}
\label{fig_link_qos}
\end{figure*}
Presenting DNS-SD is out of the scope of this paper, we only want to underline that it is possible and easy for services to broadcast their features to the other ones. We have chosen DNS-SD to demonstrate that, but other solutions may exist. 
With NSD, you are able to identify other devices on the local network that support the services on which one wants to connect. This is useful for assembling services as we propose to do so. The surrounding services are in the communication range of the device. From the device's point of view, when a service is out of range, it can no longer be used and disappears from the list of found services.
We have implemented, on the Android operating system, an application using NSD to find an image capture service (IP camera) on the near network. Android's NSD APIs simplify the effort required to implement these features. 
Figure \ref{fig_DNS_SD} shows the interface. It is divided into two parts. The top part contains the server logs and two buttons: the first starts the image capture service and the second stops it. The bottom part is for the client who searches for an image capture service. It can start and stop a discovering. When the discovering phase is on, it finds the service and collect his features (nominal/offered QoS, required memory, etc.) in a key/value form and in an eXtensible Markup Language (XML) format form.
This is the concretization of the self-defining property defined in Section \ref{sec_properties}. With NSD, services are aware of their surrounding environment. Services they found are in communication range.
In the following section, we show that, thanks to our SCC component, we are able to compute an estimation of the link QoS that will be used by our algorithm.

\subsection{Estimation of the Link QoS}\label{sec_link_qos}

In this section, we describe a method, using our SCC, for calculating the link QoS, that is the response time between services. The estimation of the link QoS is further used in our algorithm (Section \ref{sec_algorithmic_model}). Our approach allows to compute an estimation of the link QoS by using SCC inside a composition. Indeed, an SCC includes two monitors that intercept requests and responses of the functional part. These are the right places where to put a timestamp as metadata on the intercepted request and response (Figure \ref{fig_link_qos}). When the response is provided by the first SCC, the OutMonitor put a timestamp t2 on the response. When the response arrives at the second SCC, the InMonitor puts a timestamp t3 on it. So the link QoS is computed as t3-t2. This is simple, quick, consumes very little computing resources, and allows to estimate the response time between two SCC. This proposition will be used in our algorithm in the next section.
Note that the link could also be a full-fledged SCC component. The link QoS represents both a processing time for the seven-layer Open Systems Interconnection (OSI) model of computer networking \cite{art5:standardization_iso_2020} and a transport time.

In the next section, we present our algorithmic model including our core assembly algorithm.

\subsection{Algorithmic Model}\label{sec_algorithmic_model}

In Section \ref{sec_model_definition}, we define our model, introduce the terminology and notation used in the rest of the paper and present our assembling core algorithm (Section \ref{sec_core_algorithm}). In Section \ref{sec_example_of_execution}, we present an example of execution of the assembling core algorithm. The Section \ref{sec_benefits} enumerates his benefits. 

\subsubsection{Model Definition}\label{sec_model_definition}
We consider a system containing N distributed services S=\{S$_{1}$,...,S$_{N}$\}, with each service having a type  d $\in$ T=\{T$_{1}$,...,T$_{M}$\}. Services are hosted on peer nodes, each node containing one or more services. Nodes can be located anywhere and communicate with one another through a network.
Formally, a service S is a tuple (Type, QoS, Threshold), where:
\begin{itemize}
\item\textbf{S.Type} $\in$ T denotes the type of the service, its function.
\item\textbf{S.QoS} $\in$ $\mathbb{R}$ represents the service QoS nominal value (e.g. response time).
\item\textbf{S.Threshold}  $\in$ $\mathbb{N}$ represents the nominal value i.e. the threshold (number of simultaneous requests) not to exceed in order to comply with the nominal preceding defined QoS.
\end{itemize}    

A \textbf{service assembly} A is a graph A = (S,E), where E $\subseteq$ S x S is the set of bindings. Specifically, a directed edge (S$_{i}$,S$_{j}$) $\in$ E denotes that S$_{i}$ is connected to S$_{j}$. We allow multiple simultaneous bindings to the same service S by other services. The upperbound for the number of bindings to a service is the threshold value for QoS compliance (\textbf{Non-Functional constraints}). Indeed, in the worst case, services can emit simultaneously without interfering with each other to the same recipient. 
An \textbf{application template AT} is a tuple (Body, Constraints), where:
\begin{itemize}
\item \textbf{AT.Body} $\subseteq (T \times T)^p$ is a set of p values representing ordered micro-service types used to build the desired application. Specifically (t$_{i}$ ,t$_{j}$) $\in$ AT.Body denotes that services of type t$_{i}$ are connected to the services of type t$_{j}$.
\item \textbf{AT.Constraints} $\in$ ($\mathbb{N}^{*}$ $\cup$ \{{$\bowtie$}\})$^{p}$ is a set of p values of $\mathbb{N}^{*}$ $\cup$ \{{$\bowtie$}\} representing the number of services that should be connected according to each tuple (t$_{i}$,t$_{j}$) $\in$ AT.Body. $\mathbb{N}^{*}$ = $\mathbb{N}$ $\setminus$ \{0\} is the set of non-zero natural numbers. Specifically: b$_{k}$ = (t$_{i}$,t$_{j}$) $\in$ AT.Body and c$_{k}$ $\in$ AT.Constraints means that all services of type t$_{i}$ should be connected to c$_{k}$ services of type t$_{j}$. $\bowtie$ means that services of type t$_{i}$ should be connected to all available services of type t$_{j}$. These are \textbf{structural constraints}.
\end{itemize} 

A service assembly A is \textbf{AT compliant} if A is compliant with the application template i.e. services are connected according to AT.Body. 

A service assembly A is \textbf{ATS compliant} (S for structural constraints) if A is AT compliant with respect to AT.Constraints (\textbf{structural constraints}). 

A service assembly A is \textbf{ATSF compliant} (SF for structural and non-functional constraints) if A is ATS compliant and the number of bindings to each service is upper bounded by the threshold value (\textbf{Non-Functional constraints}). 

For a given template AT, there may be more than one AT compliant assembly.

Given S$_{k}$ $\in$  S with S$_{k}$ of type t$_{k}$, S$_{k}$ is a \textbf{starting service} if there is no tuple (t$_{i}$,t$_{j}$) $\in$ AT.Body where t$_{j}$=t$_{k}$.

Given S$_{k}$ $\in$  S with S$_{k}$ of type t$_{k}$, S$_{k}$ is an \textbf{ending service} if there is no tuple (t$_{i}$,t$_{j}$) $\in$ AT.Body where t$_{i}$=t$_{k}$.

For an application template AT, in AT.Body, we state that there is only one type of starting service. 
A second starting type means a second application, so a second template.

For a given AT compliant service assembly A and a starting service S$_{i}$ $\in$ S of type t$_{i}$, we define a subgraph p$_{Si}$ $\subseteq$ A, formed by all graphs starting from node S$_{i}$. 

\begin{figure}[!t]
\centering
\includegraphics[width=0.5\textwidth]{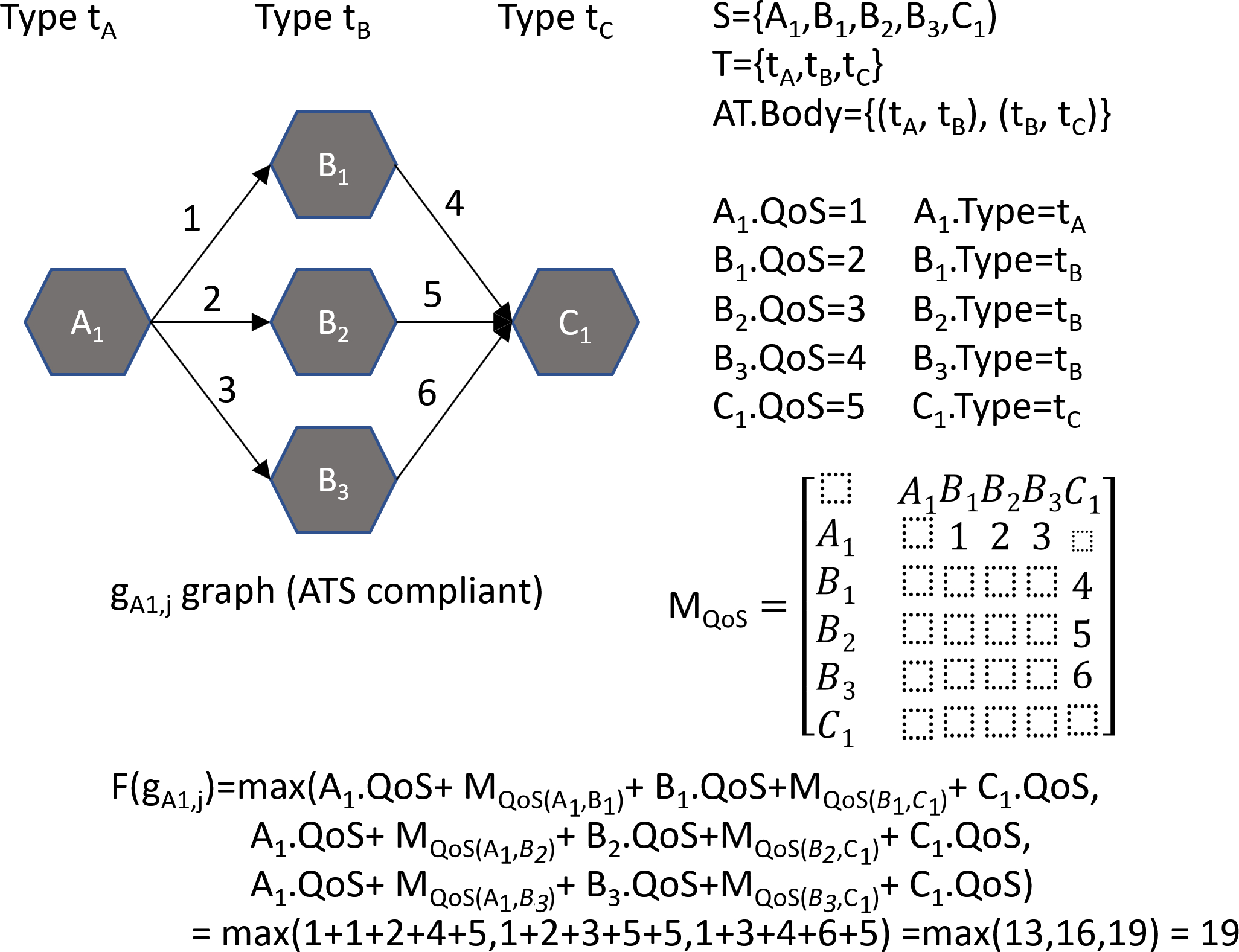}
\caption{F(g$_{A1,j}$) calculation example.}
\label{fig_calculation_example}
\end{figure}

For a subgraph g$_{Si,j}$ of p$_{Si}$, we define F(g$_{Si,j}$) as Max (processing time from the starting service S$_{i}$ to an ending service). Processing time is computed as the sum of transfer times between services (computed link QoS, see algorithm 1) and processing times (nominal S.QoS) of each service composing the path (Figure \ref{fig_calculation_example}). We choose the worst case, i.e. the maximum of the processing times.
We try, in our algorithms, to choose the best organizations that minimize the set of F(g$_{Si,j}$) i.e. the maximum processing time of all subgraphs g$_{Si,j}$.

Since our services are stateless, they do not maintain the interaction state between service invocations, i.e., a client request is served in complete isolation, without relying on information from previous requests. Hence, we assume that the state of the interaction between the client and the service is kept on the user's side, and requests include all information necessary for their processing. Service statelessness enhances (i) decoupling of interacting services (ii) flexibility of the model, since it allows for easily rearranging the assembly at run time and, (iii) scalability.
The table \ref{tab_symbols_used} summarizes the notation used in this paper. The table also includes additional symbols which will be introduced in the next sections.

\begin{table}
\small
\caption{Symbols Used in this Paper}
\label{tab_symbols_used}       
\begin{tabular}{ll}
\hline\noalign{\smallskip}
symbols & definitions \\
\noalign{\smallskip}\hline\noalign{\smallskip}
N & Number of services (peers) \\
S & Set of services S = \{S$_{1}$,...,S$_{N}$\} \\
M & Number of types \\
T & Set of service types T = \{t$_{1}$,...,t$_{M}$\} \\
A & Service Assembly A=(S,E) \\
S.Type & Service type of S \\
S.Threshold & Number of services maximum bindings \\
p$_{Si}$ & Subgraph with S$_{i}$ starting service \\
g$_{Si,j}$ & Subgraph j of p$_{Si}$ ATS compliant \\
AT.Body & Application body template \\
AT.Constraints & Structural constraints \\
S.QoS & QoS nominal value \\
F(g$_{Si,j}$) & Maximum processing time of the subgraph g$_{Si,j}$ \\
G$_{AT}$ & AT compliant graph \\
M$_{QoS}$ & Matrix of all computed link QoS (time transfer) \\
\noalign{\smallskip}\hline
\end{tabular}
\end{table}

\subsubsection{Assembling Core Algorithm}\label{sec_core_algorithm}

The assembling algorithm is structured by three sub-algorithms, here noted 1 to 3.

Algorithm \ref{alg1} build an \textbf{AT compliant} graph and compute all link QoS.
We call G$_{AT}$ this graph and M$_{QoS}$ $\in$ M$_{N,N}$($\mathbb{R}$) the matrix of all computed link QoS. 

\begin{algorithm}
\caption{AT compliant}
\label{alg1}
\begin{algorithmic}[1]
\FORALL{S$_{i}$ starting service of type t$_{i}$}
\FORALL{(t$_{i}$,t$_{j}$) $\in$ AT.Body:}
\STATE S$_{i}$ sends a request to all services of type t$_{j}$.
\ENDFOR
\ENDFOR
\STATE 
\IF{a service S$_{i}$ service of type t$_{i}$, receives a request}
\STATE it computes the QoS of the link between him and the sender.
\FORALL{(t$_{i}$ , t$_{j}$) $\in$ AT.Body:}
\STATE S$_{i}$ sends a request to all services of type t$_{j}$.
\ENDFOR
\ENDIF
\end{algorithmic}
\end{algorithm}
We will now take into account AT.Constraints (Algorithm \ref{alg2}). AT.Constraints introduces restrictions in G$_{AT}$ so that the resulting graph will be a subset of G$_{AT}$. Taking into account AT.Constraints is equivalent to make a choice among available paths. Restriction can be of two types for a path: k $\in$ N$^{*}$ or $\bowtie$. k means that we have to choose k services among services of the same type. $\bowtie$ means that we choose all available services of the specified type. Algorithm \ref{alg1} builds G$_{AT}$ as if there are no restrictions ($\bowtie$ case).
We sort all subgraphs g$_{Si,j}$ of p$_{Si}$ in F(g$_{Si,j}$) ascending order. Note that this is not necessary if we don't want to minimize the set of F(g$_{Si,j}$) but only to find a suitable assembly. Note that if there were no non-functional constraints, the first combination would be optimal since all g$_{Si,j}$ are sorted ascending.

\begin{algorithm}
\caption{ATS compliant}
\label{alg2}
\begin{algorithmic}[1]
\FORALL{p$_{Si}$ of G$_{AT}$ with S$_{i}$ starting service}
\STATE We compute all subgraphs g$_{Si,j}$ of p$_{Si}$ taking into account AT.Constraints.
\STATE Sort of all subgraphs g$_{Si,j}$ in F(g$_{Si,j}$) ascending order.
\ENDFOR
\end{algorithmic}
\end{algorithm}
Algorithm \ref{alg3} checks if functional constraints are fulfilled and will find a suitable combination if it exists but will not be necessarily the optimal in terms of min(F(g$_{Si,j}$)).
Indeed, to avoid time-consuming tasks, we are not aiming at the optimal value.
The combination will, however, be near the optimal solution since g$_{Si,j}$ are sorted in ascending order as specified in algorithm \ref{alg2}. 

\begin{algorithm}
\caption{ATSF compliant}
\label{alg3}
\begin{algorithmic}[1]
\STATE We choose one subgraph g$_{Si,j}$ for each p$_{Si}$ taking into account AT.Constraints.
\FORALL{combination}
\IF{number of bindings to each service is less or equal than the threshold value (S.Threshold)}
\STATE This combination has been built according to the application template AT in respect of structural (number of bindings) and non-functional constraints (QoS compliance).
\STATE \textbf{Break}
\ELSE
\STATE The combination is rejected.
\ENDIF
\ENDFOR
\end{algorithmic}
\end{algorithm}

\subsubsection{Example of Execution of the Assembling Core Algorithm}\label{sec_example_of_execution}

We present in this section an example of execution of the assembling core algorithm. It includes seven services named A$_{1}$, A$_{2}$, A$_{3}$ of type t$_{A}$, B$_{1}$, B$_{2}$, B$_{3}$ of type t$_{B}$ and C$_{1}$ of type t$_{C}$.\\
\\
S=\{A$_{1}$,A$_{2}$,A$_{3}$,B$_{1}$,B$_{2}$,B$_{3}$,C$_{1}$\}\\
T=\{t$_{A}$,t$_{B}$,t$_{C}$\}\\
A$_{1}$.Type=t$_{A}$\\
A$_{2}$.Type=t$_{A}$\\
A$_{3}$.Type=t$_{A}$\\
B$_{1}$.Type=t$_{B}$\\
B$_{2}$.Type=t$_{B}$\\
B$_{3}$.Type=t$_{B}$\\
C$_{1}$.Type=t$_{C}$\\

These are the service QoS nominal values (e.g. response time) obtained with a calibration procedure. Specified thresholds (number of simultaneous requests) are the values not to exceed in order to comply with the defined nominal QoS.\\
\\
A$_{1}$.QoS=1, A$_{1}$.Threshold=1\\
A$_{2}$.QoS=1, A$_{2}$.Threshold=1\\
A$_{3}$.QoS=1, A$_{3}$.Threshold=1\\
B$_{1}$.QoS=2, B$_{1}$.Threshold=2\\
B$_{2}$.QoS=3, B$_{2}$.Threshold=3\\
B$_{3}$.QoS=4, B$_{3}$.Threshold=1\\
C$_{1}$.QoS=5, C$_{1}$.Threshold=3\\

All services should be assembled according to the following application template AT:\\
\\
AT.Body=\{(t$_{A}$,t$_{B}$), (t$_{B}$,t$_{C}$)\}\\
AT.Constraints=\{2,1\}

The AT template means that the t$_{A}$ services should be connected to two t$_{B}$ services and all t$_{B}$ services should be connected to one t$_{C}$ service.

\begin{figure*}[!t]
\centering
\includegraphics[width=1.0\textwidth]{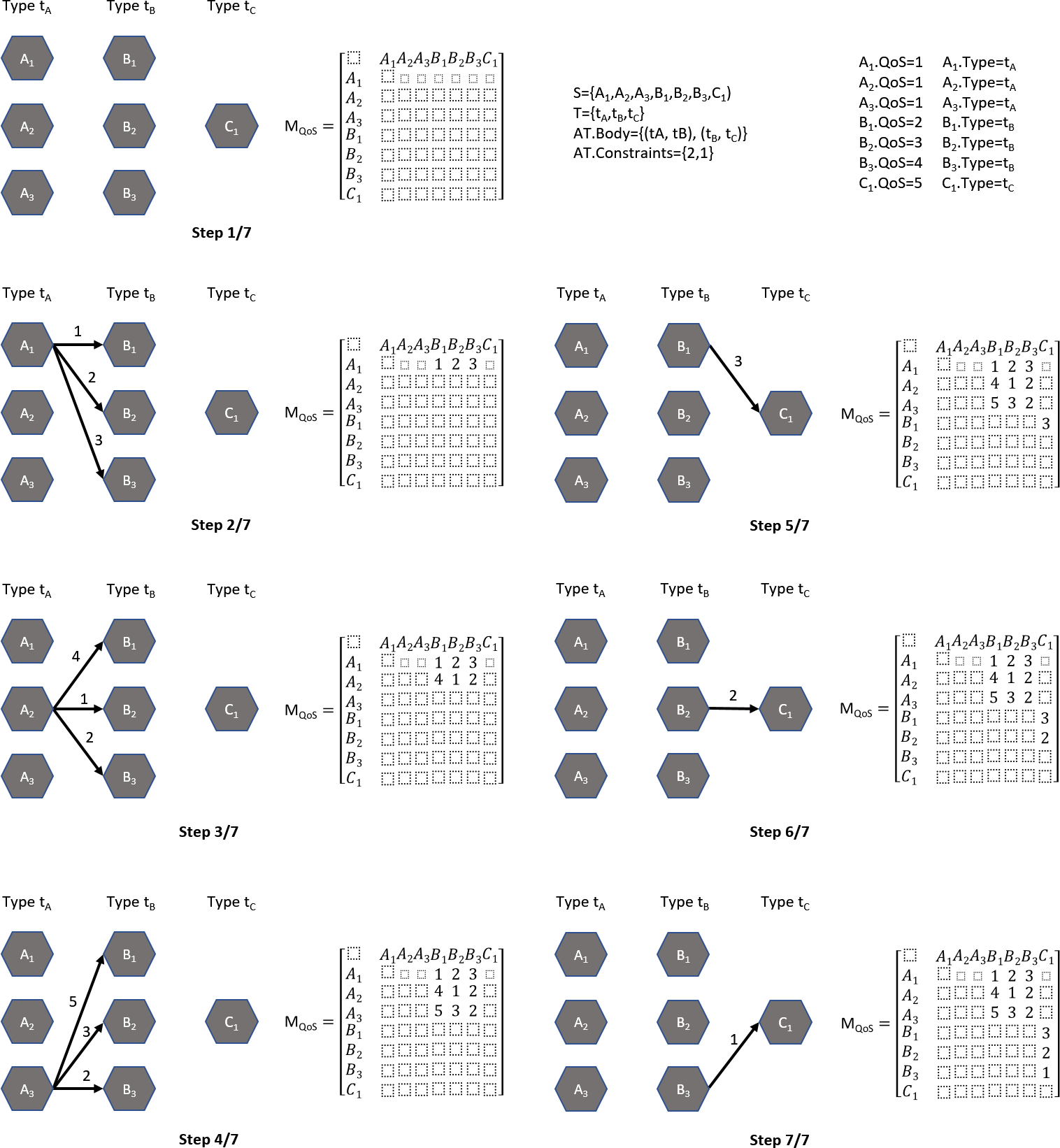}
\caption{Algorithm 1 - AT compliant graph / M$_{QoS}$ design.}
\label{fig_algo1}
\end{figure*}

\begin{figure}[!t]
\centering
\includegraphics[width=0.26\textwidth]{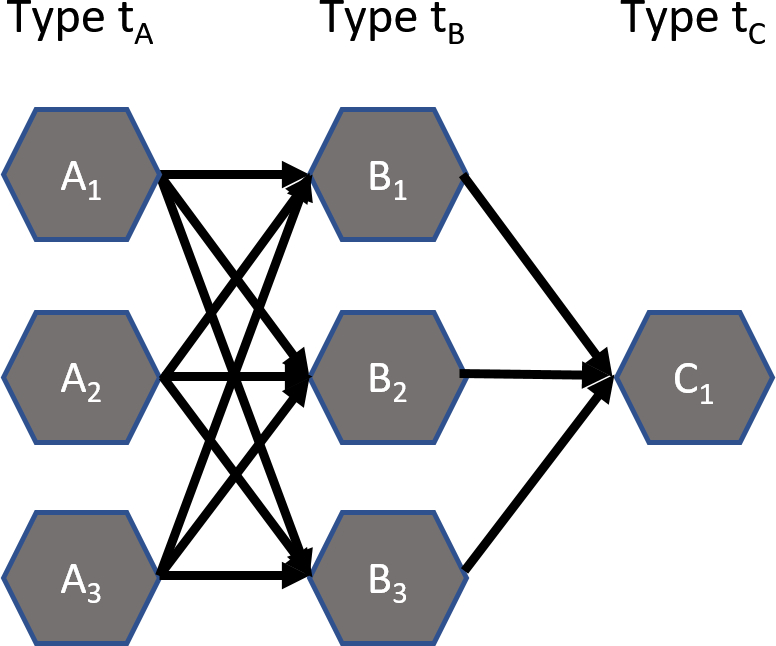}
\caption{AT compliant graph: G$_{AT}$.}
\label{fig_at_compliant_graph}
\end{figure}

\textbf{Algorithm 1 execution:} We applied the Algorithm 1 to determine the M$_{QoS}$ matrix including all computed link QoS.
Services send a request according to the AT template (See Figure \ref{fig_algo1}). Step 1 shows the initial state of services. Each service is aware of the existence of others. According to AT template, t$_{A}$ services should be connected to t$_{B}$ services. So, t$_{A}$ services send a request to t$_{B}$ services. Step 2 shows the service A1 sending a request to B$_{1}$, B$_{2}$, and B$_{3}$. B$_{1}$, B$_{2}$, and B$_{3}$ compute the link QoS (response time) and inscribe it in M$_{QoS}$. The same procedure is repeated for each of the first four steps. According to AT template, t$_{B}$ services should be connected to t$_{C}$ service. So, t$_{B}$ services send a request to t$_{C}$ services (steps 5 to 7). Algorithm 1 build an AT compliant graph called G$_{AT}$ (Figure \ref{fig_at_compliant_graph}) and compute all link QoS.

\begin{figure*}[!t]
\centering
\includegraphics[width=1.0\textwidth]{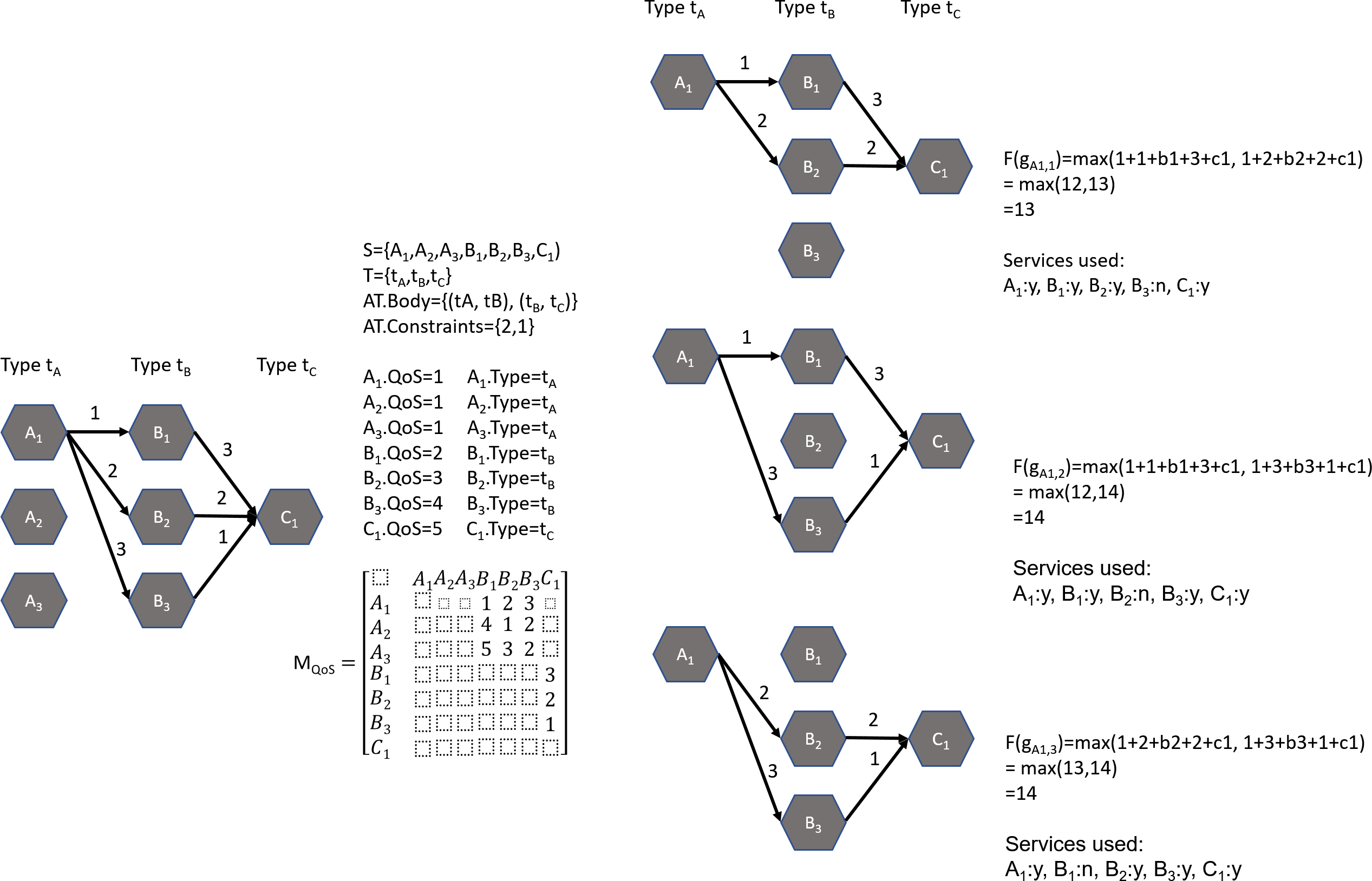}
\caption{Algorithm 2 - ATS compliant graph: Computing of all subgraphs g$_{A1,j}$ of p$_{A1}$ taking into account AT.Constraints.}
\label{fig_algo2_a1}
\end{figure*}

\begin{figure*}[!t]
\centering
\includegraphics[width=1.0\textwidth]{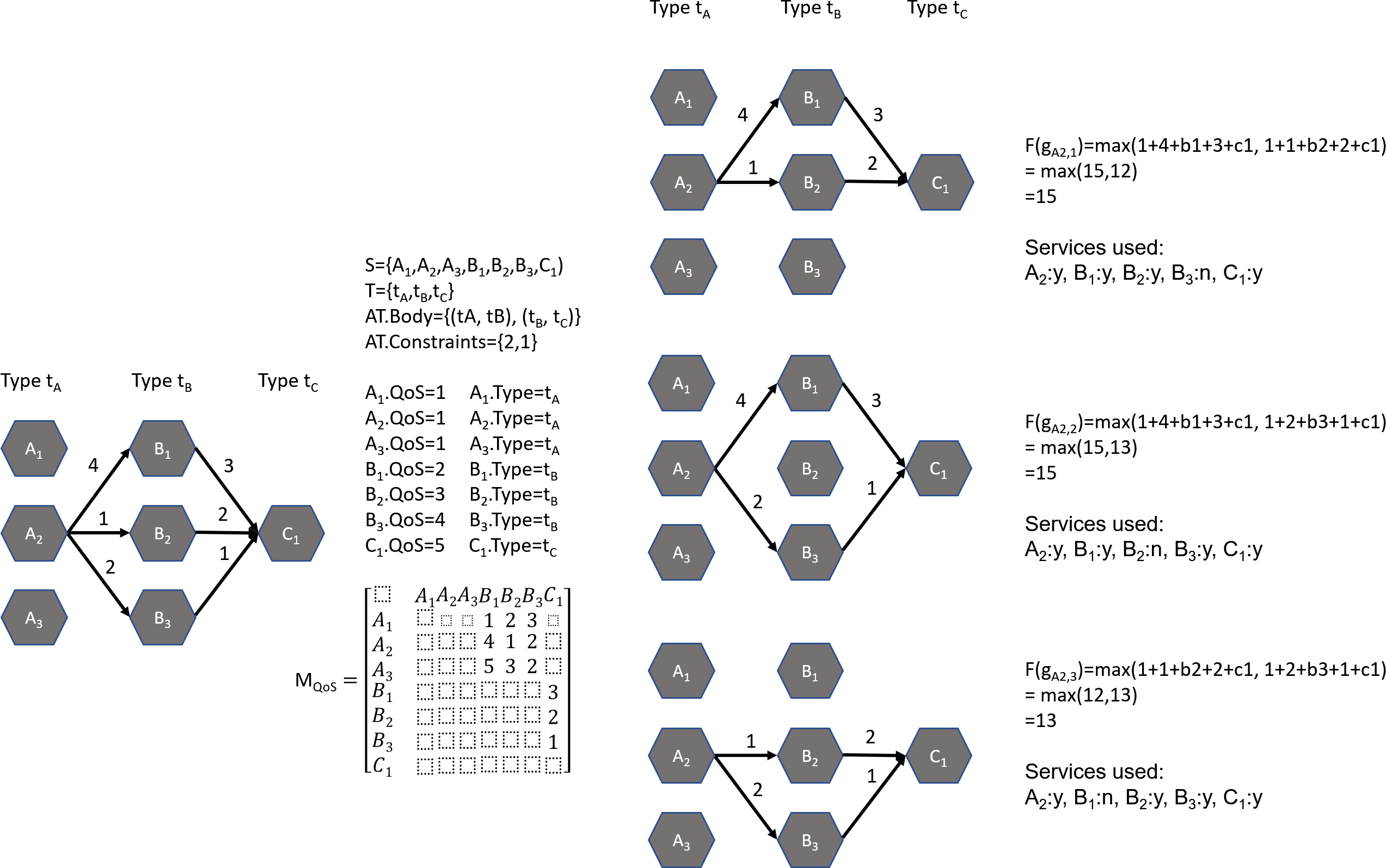}
\caption{Algorithm 2 - ATS compliant graph: Computing of all subgraphs g$_{A2,j}$ of p$_{A2}$ taking into account AT.Constraints.}
\label{fig_algo2_a2}
\end{figure*}

\begin{figure*}[!t]
\centering
\includegraphics[width=1.0\textwidth]{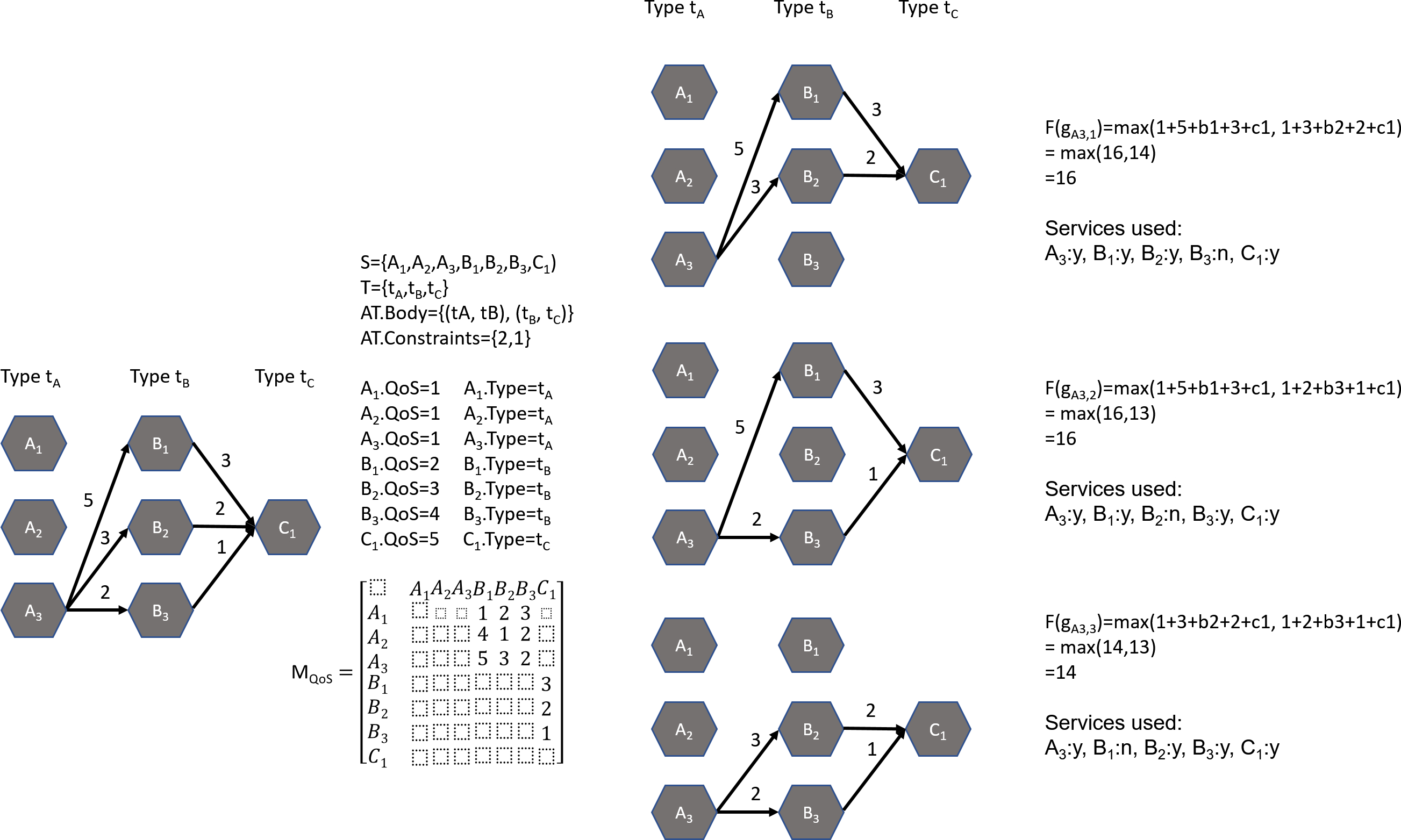}
\caption{Algorithm 2 - ATS compliant graph: Computing of all subgraphs g$_{A3,j}$ of p$_{A3}$ taking into account AT.Constraints.}
\label{fig_algo2_a3}
\end{figure*}

\begin{figure}[!t]
\centering
\includegraphics[width=0.29\textwidth]{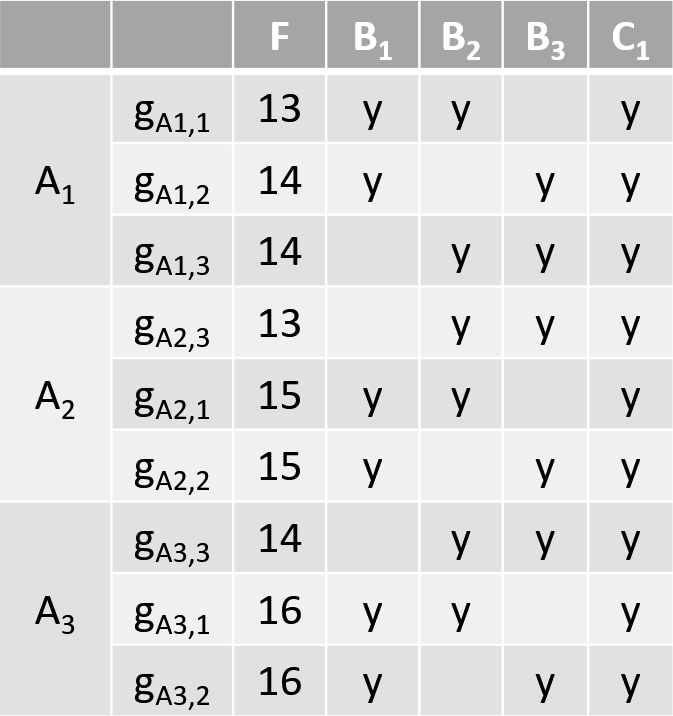}
\caption{Results of the algorithm 2 (y means that the corresponding service is used).}
\label{fig_results_of_algo2}
\end{figure}

\textbf{Algorithm 2 execution:} We take into account the structural constraints AT.Constraints=\{2,1\} by applying the algorithm 2. There are three subgraphs to consider, each with a different starting service: p$_{A1}$, p$_{A2}$, and p$_{A3}$. For each of them, we determine all subgraphs taking into account AT.Constraints i.e. t$_{A}$ services should be connected to two t$_{B}$ services and all t$_{B}$ services should be connected to one t$_{C}$ service.
For each subgraph s, we compute F(s) i.e. the maximum processing time of the subgraph.
Figure \ref{fig_algo2_a1} deals with p$_{A1}$, Figure \ref{fig_algo2_a2} deals with p$_{A2}$, and Figure \ref{fig_algo2_a3} deals with p$_{A3}$. We also indicate which services are used for each of the subgraph. Figure \ref{fig_results_of_algo2} shows results sorted by ascending order.

\begin{figure*}[!t]
\centering
\includegraphics[width=1.0\textwidth]{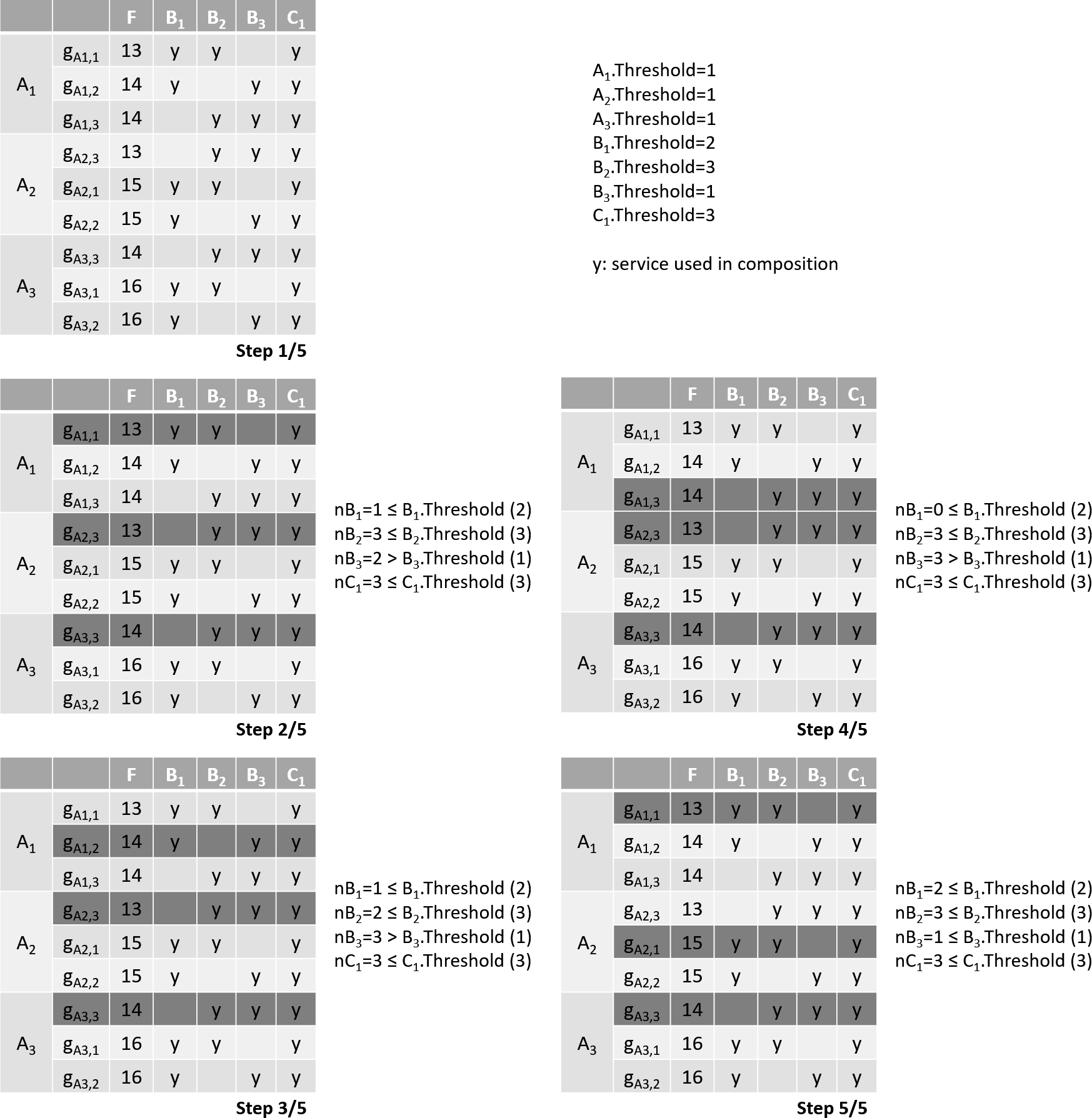}
\caption{Algorithm 3 - ATSF compliant graph (The chosen graphs are colored in dark gray).}
\label{fig_algo3}
\end{figure*}

\begin{figure}[!t]
\centering
\includegraphics[width=0.26\textwidth]{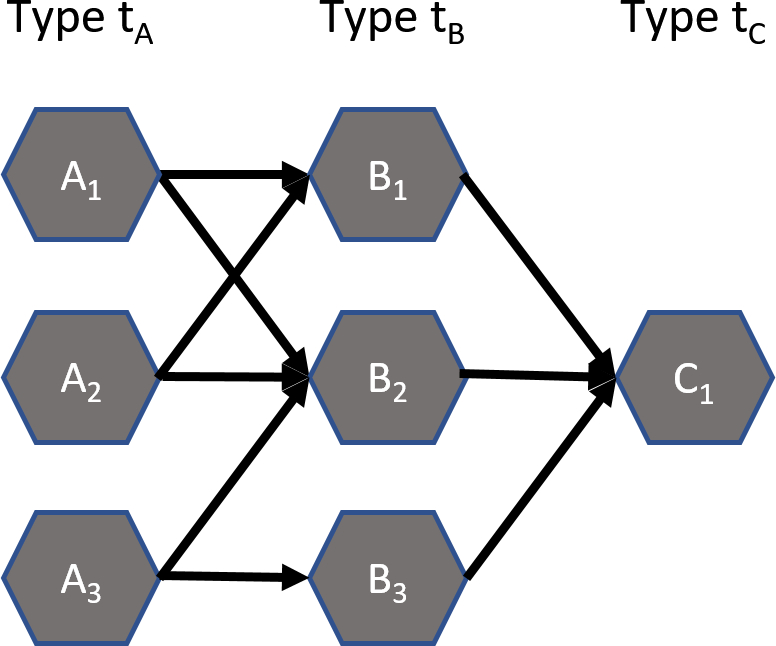}
\caption{Final suitable assembly.}
\label{fig_final_assembling}
\end{figure}

\textbf{Algorithm 3 execution: }For each combination, choosing one subgraph for each starting service (A$_{1}$, A$_{2}$, A$_{3}$), we check if non-functional constraints are fulfilled i.e. if the number of bindings to each service is less than or equal to the threshold value (S.Threshold). Otherwise, the combination is rejected (Figure \ref{fig_algo3}). We compute the number of used B$_{1}$, B$_{2}$, B$_{3}$, and C$_{1}$ service for each combination. After five steps, the algorithm finds a suitable combination. The final assembly is given in Figure \ref{fig_final_assembling}.

\subsubsection{Benefits of the Assembling Core Algorithm}\label{sec_benefits}

The assembling core algorithm can potentially be run at anytime:
\begin{itemize}
\item When a service appears, or disappears on the network. A new service can be more efficient (better QoS) or can have a better link. 
\item When a service has failed and should be replaced. The assembling can thus be considered as a kind of adaptation mechanism. The algorithm avoids the service and proposes a new assembling. The composition is more robust and reliable.
\item When communication (link) is becoming bad. For example, when a service moves away from another one or in case of electromagnetic disturbed environment. A too bad communication can be the cause of the disappearance of a service.
\end{itemize}  
The assembling core algorithm can thus potentially be used as self-adjusting, self-adapting, and self-healing mechanisms.

For the user, it simplifies fastidious manual assembling of a lot of devices. The service composition is performed automatically without human intervention.

The assembling is dynamic. The application template AT is provided to perform a function. We can change later the template and thus change the function by reusing some or all of the current services.

Some assembling results lead to well-known organizations like Edge \cite{art5:jain_fog_2016,art5:ahmed_mobile_2017}, Dew \cite{art5:wang_definition_2016} or Fog computing\cite{art5:datta_fog_2015,art5:sarkar_assessment_2015} \cite{art5:skala_scalable_2015}. 

The algorithm can be useful for the run time, as mentioned, but also for the designing of the whole application. The architect has to make choices like the number of services of each type, in particular, to prepare and deploy in the environment so that a future need is answered. He can use the algorithm to simulate different organizations before deploying and implementing them in the real world (Self-simulation property).

\subsection{Resource Consumption and Limitations of the Algorithm}\label{sec_resource_consumption}

In this section, we estimate the limitations of the algorithm on a real IoT platform. The aim of this section is to estimate the maximum number of services, SCC in our case, that could be assembled with our algorithm in real conditions, in a reasonable amount of time and, if possible, low-resource consumption (memory). In Section \ref{sec_equipement}, we present the equipment used for the implementation of our algorithm. In Section \ref{sec_combination_analysis}, we determine the worst case for assembling constraints: when the algorithm might experience difficulties. Finally, in Section \ref{sec_layouts_and_result_analysis}, we present the chosen layouts and we analyse the experimental results.

\subsubsection{The Equipment Used for the Implementation}\label{sec_equipement}

The raspberry pi has been chosen to make our experimentations. Our experimental devices 
are composed of:
\begin{itemize}
\item \textbf{Raspberry Pi card}, proposed by the British Raspberry Pi Foundation, is a one-board nano-computer, about the size of an ARM processor-based credit card \cite{art5:raspberry_2017}. 
\item \textbf{Android Things operating system} is an embedded operating system based on Android designed by Google \cite{android_things}. This system is meant to be used on devices linked to the Internet of Things. It is therefore designed to use as little memory as possible and to be energy-efficient.
\item \textbf{3.5-inch touch-screen monitor}, compatible with the Raspberry Pi card, allows users to create all-in-one projects such as tablets, entertainment systems, embedded projects and devices for the Internet of things using interaction with the touch screen \cite{raspberry_display}.
\end{itemize}  
 
\subsubsection{Combination Analysis}\label{sec_combination_analysis}

Constrains are defined for each level. They represent the number of services that should be connected. The architect has to specify the number of services that should be chosen among available services of the same type. 
At each level, k services have to be chosen among n available services. The number of combinations is 
\begin{equation}
\binom{n}{k}=C_{n}^{k}=\frac{n!}{k!(n-k)!}
\end{equation}
$\binom{n}{k}$ is read as "n choose k".
There are a few combinations if k is very small or very big but become maximal if k is close to n/2. 
Choosing the case of n/2 services should be avoided and is extremely rare in real life. Common cases consist to choose only one service (1 of n) or all available services (n of n).

Notice that our study seems to be similar to the shortest path problem which can be solved by several well-known algorithms (Dijkstra's \cite{lacorte_analysis_2018}, Bellman-Ford \cite{schambers_route_2018}, A* search \cite{lacorte_analysis_2018}, Floyd-Warshall \cite{ramadhan_prim_2018}, Johnson \cite{anitha_network_2018}, or Viterbi \cite{zhou_efficient_2018}) but is very different and more complex because (i) starting and ending services are not unique, (ii) they are searching for a path while we are searching for a sub-graph and (iii) the number of bindings to a node is upper bounded.

Usually, an application is monolithic, i.e. the same for each user.
It offers features to meet a set of needs even if the user does not use them all.
With our approach, the performance of the application can only be better because the assembly is dynamic.
Customized applications are created according to the user's current needs.
A service is integrated into our assembly only if necessary.
 
\subsubsection{Layouts and Result Analysis}\label{sec_layouts_and_result_analysis}

\begin{figure*}[!t]
\centering
\includegraphics[width=1.0\textwidth]{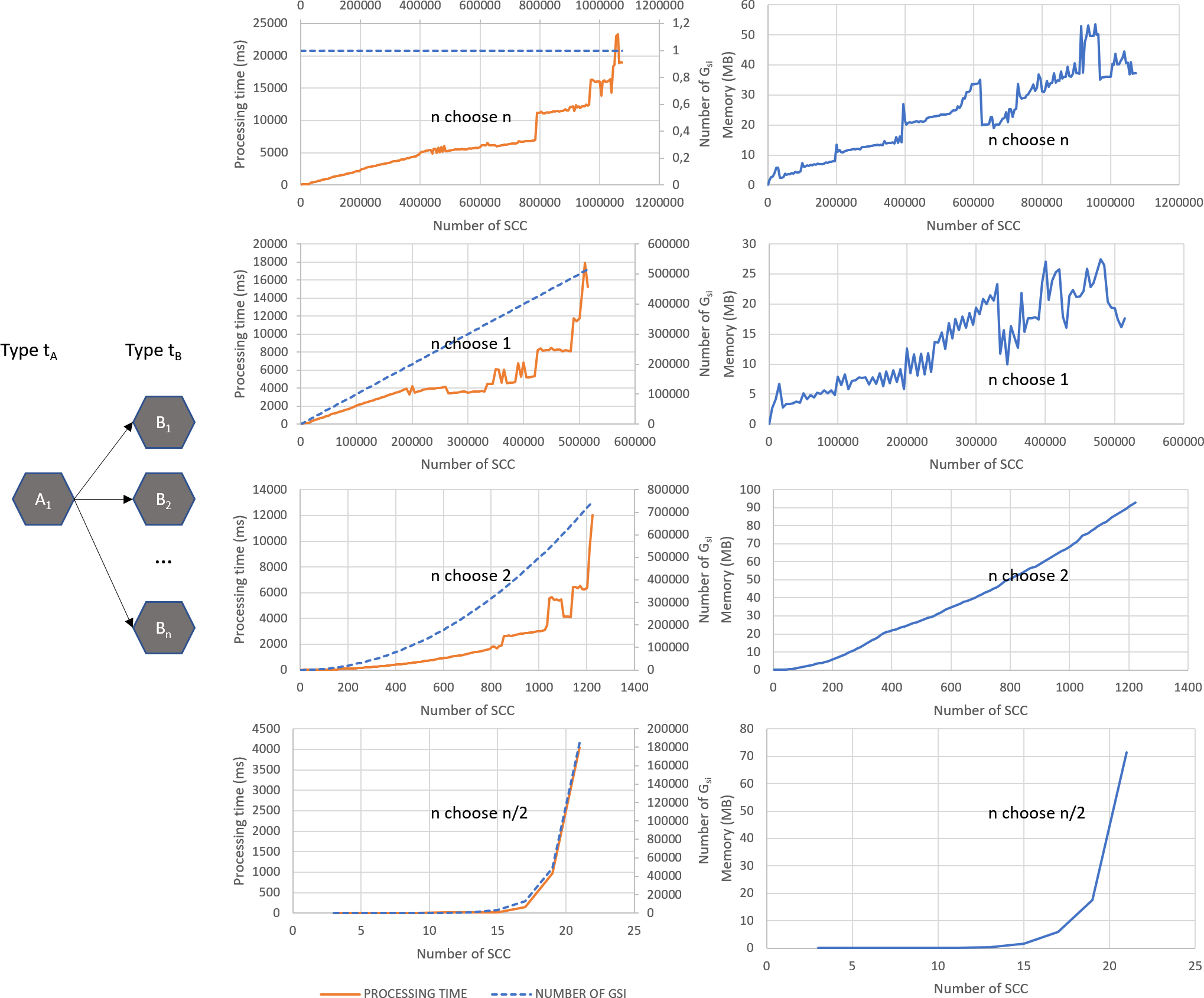}
\caption{One-layer layout and results of its implementation.}
\label{fig_one_layer_layout}
\end{figure*}

We have chosen two services layouts. 
Threshold values, link QoS and SCC processing time are random.
The first layout has only one level. It is the simplest one and allows us to evaluate the maximum number of SCC able to be assembled for a given layer. This arrangement is made to evaluate the limitation of chosen all, 1, 2, or n/2 services. The scheme of the layout is given in Figure \ref{fig_one_layer_layout}.
If the architect asks to choose all available services or only one of them, which are the most commonly used requirements, the algorithm is able to assemble respectively 1 million and 500 000 services.
If the demand becomes more combinatorial like choosing 2 or the worst n/2 services among all available SCC of a given type, the number of assembled SCC decreases respectively to 1200 and 20.
The amount of memory necessary for the assembling is reasonable and doesn't exceed 100 MB in all cases.
The number of g$_{Si,j}$ computed by the algorithm is also given for information for all four cases.
The processing time does not exceed 20s in all cases. It might become, however, excessive depending on the case study but can easily be reduced if we accept to reduce the size of the assembling.

\begin{figure*}[!t]
\centering
\includegraphics[width=1.0\textwidth]{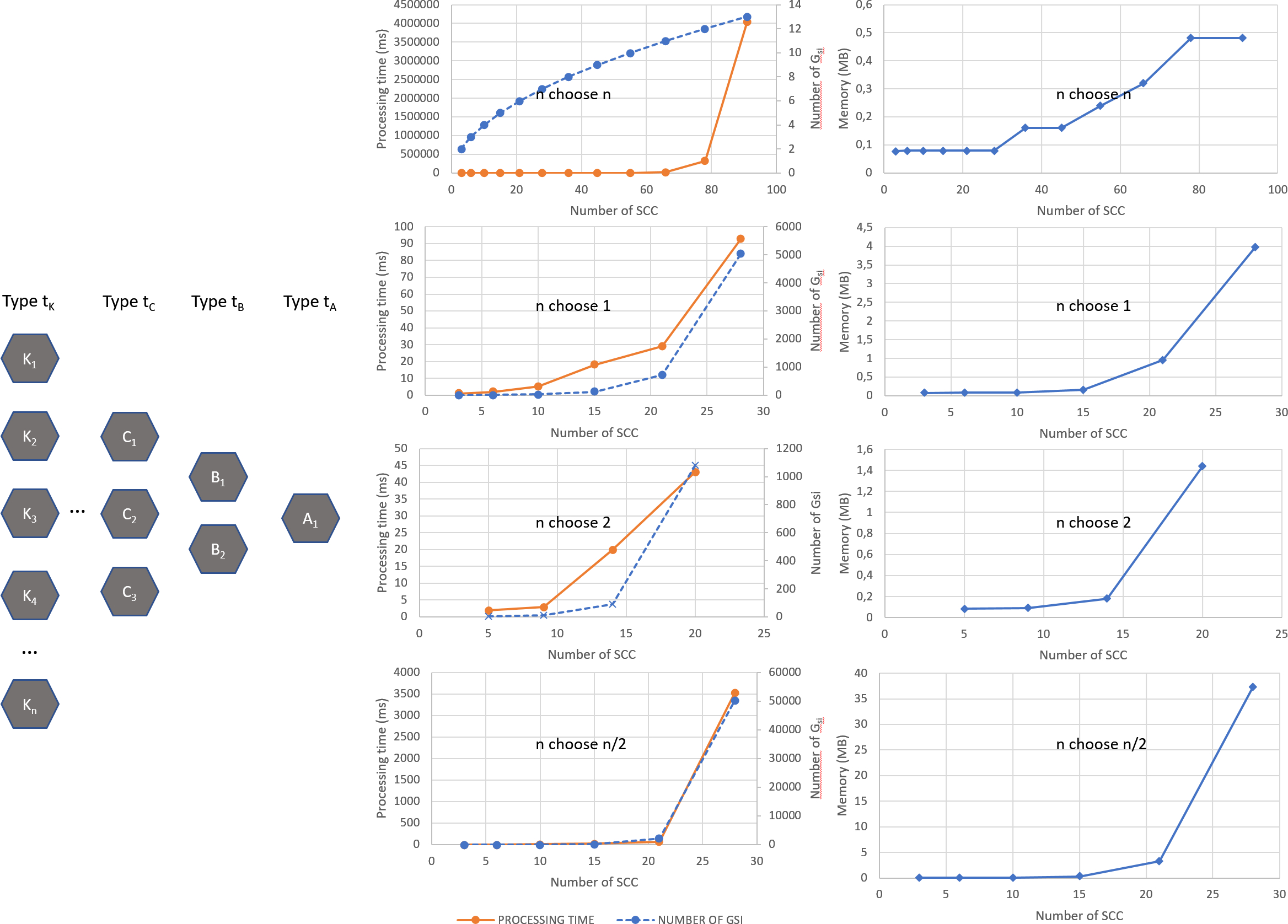}
\caption{Pyramidal decreasing layout and results of its implementation.}
\label{fig_pyramidal_decreasing_layout}
\end{figure*}

The second layout is excessively combinatory and is designed to push the algorithm as far as we can to its very limits.
This layout consists in building a pyramidal organization by decreasing the number of SCC of one at each level until the raspberry pi is at the end (Figure \ref{fig_pyramidal_decreasing_layout}).
If the architect asks to choose all available services or only one among them, the algorithm is able to assemble respectively 91 and 28 SCCs.
If the demand consists in choosing 2 or the n/2 services among all available SCC of a given type, the number of assembled SCC decreases respectively to 20 and 28.
The amount of memory necessary for the assembling remains reasonable and doesn't exceed 37 MB in all cases.
The processing time is reasonable and doesn't exceed 3.5s for all the last three cases (n choose 1, n choose 2 and n choose n/2) but becomes excessive if we choose all services at each level (n choose n) especially with 91 SCCs. This remains, however, correct for 55 SCCs (9 layers, 10 types), the processing time being 2.3s.

Measurements are taken in a way that the measuring process has none or few effects on the evaluated service.
To minimize the impact of collecting measurement data, the QOsControl component is a thread that regularly requests their measurements to the two monitors and write them in a log file which is collected at the end of the experiment.

Analysis of the approximate resource overhead (computation and memory) has been written in \cite{lemoine_iot_2020}.
We have analyzed the extra code needed by the sub-components located in the SCC membrane: InMonitor, OutMonitor and QoSControl. 
We have used profiling tools to measure memory and CPU consumption of our approach at runtime.
The memory and CPU consumption of our control mechanism is very low for common usage. However, it must be compared to the size of the functional part. The control mechanism should have no or a little influence on the monitored service, so it must represent a low percentage of resource usage compared to it. Indeed, if the micro-service is too small, the processing done by the control mechanism can be in the same order of magnitude.

In conclusion, as expected, increasing the number of layers (the number of service types) increases the number of combinations. The number of layers should not exceed 9 (10 different service types) to obtain a processing time of a few seconds whatever the layout and constraints.
The amount of used memory (100 MB in the worst case) is correct for a Raspberry Pi 3 Model B that has 1 GB available for all organizations.
The number of assembled SCC can be higher (up to 1 million) depending on the chosen services arrangement and constraints. 
As we said, n choose 1 case and n choose n cases are commonly used and will present good results. In addition, computations could be distributed on all devices, which could significantly improve performance. The higher the number of SCCs, the greater the complexity, but the higher processing capabilities. Algorithms 1 and 2 can be easily parallelized since each startup service could do a part of the work. More complex distributions are possible.

\section{Case Study: Medical Warning System}\label{sec_case_studies}

In this section, we deal with a case study related to the Internet of Things where our approach could prove useful. 
The idea is to show that our algorithm can be applied easily to use practical cases and can propose a suitable dynamically service assembly with a good processing time.

\begin{figure}[!t]
\centering
\includegraphics[width=0.23\textwidth]{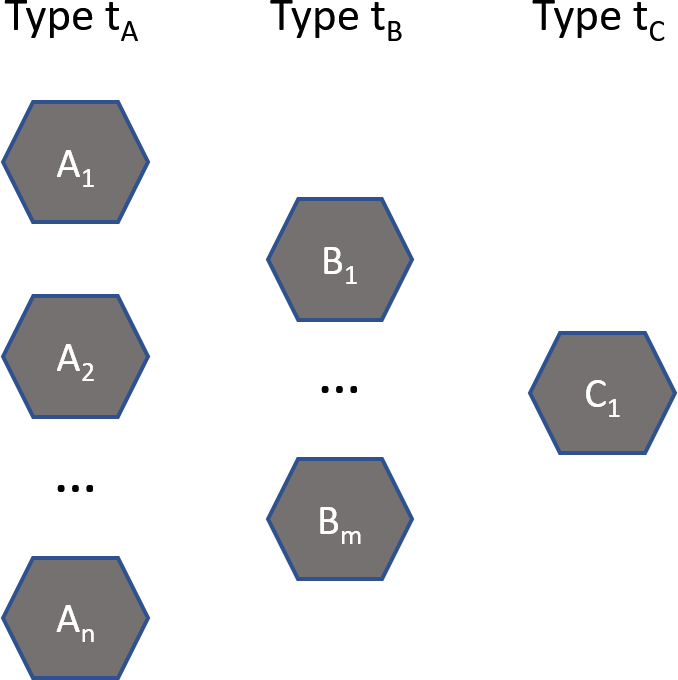}
\caption{Three layers pyramidal decreasing layout.}
\label{fig_three_layers}
\end{figure} 

\begin{figure*}[!t]
\centering
\includegraphics[width=1.0\textwidth]{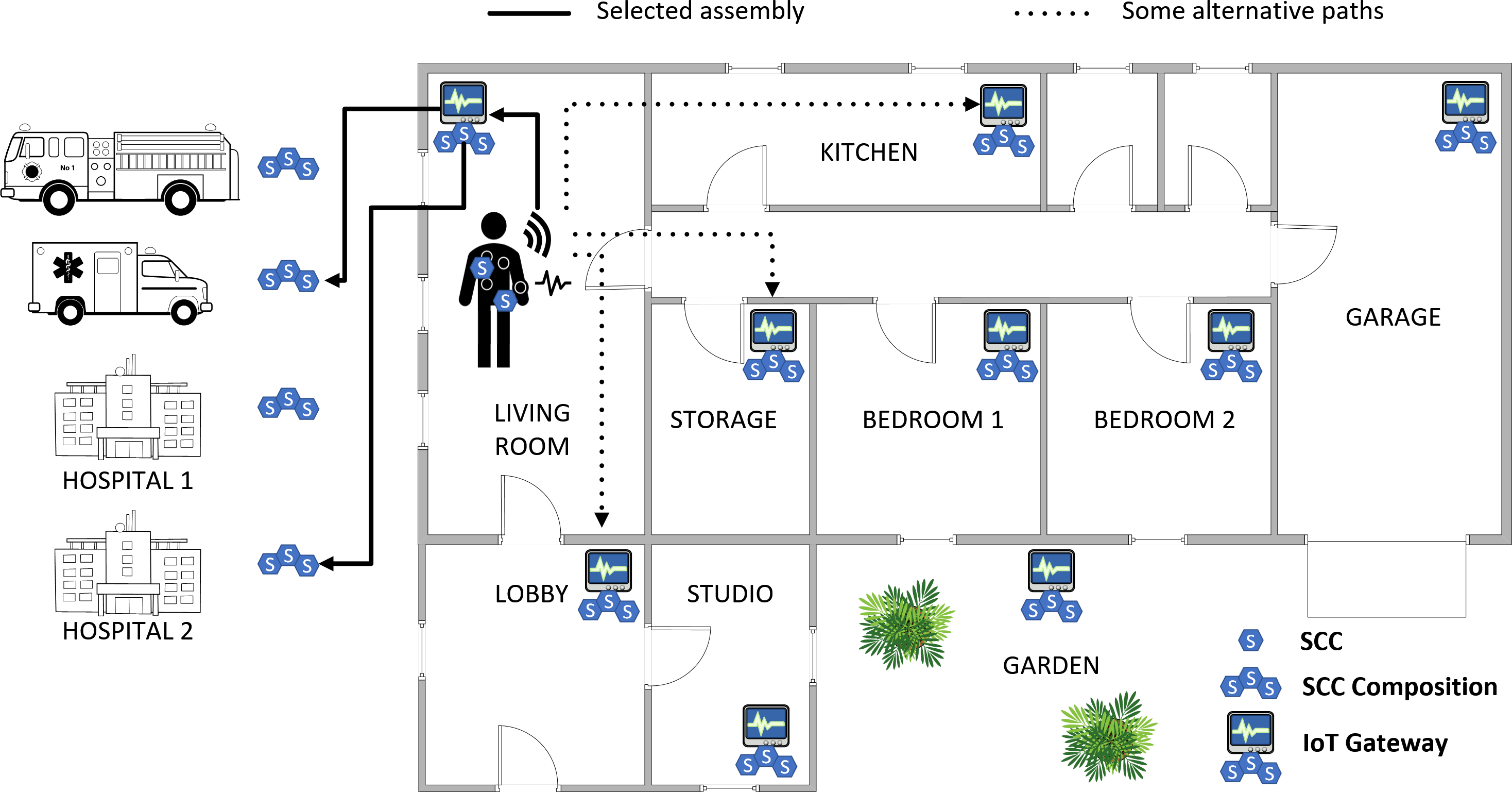}
\caption{Self-assembly for a medical warning system.}
\label{fig_usecase2_house_plan}
\end{figure*}

Due to the importance of observing the medical state of patients who are suffering from acute diseases, especially cardiovascular diseases, a continuous remote patient monitoring is essential. The monitoring system can provide real-time analytical services, prediction and alerts in case of an emergency for users carrying out any activity, anywhere and anytime. It can recognize an abnormal user activity and can thereby detect falls on the ground or sickness. The user and the associated medical staff receive local notification in case of current or near-future predicted emergency. In addition, it reduces the response time of the system and increased his reliability.
With the help of wearable wireless sensors, the system provides a continual access to medical parameters of a patient. IoT Gateways are located in every room in the house in a way to follow the patient. User sensors are always close to a gateway.  They are equipped with computational capacity. They monitor the current state of the patient and provide a means to predict future medical condition via machine learning methods and artificial intelligence algorithms. They include the following capabilities: collect and aggregate data from the available sensors, implement data analysis and extract valuable information and knowledge from the incoming raw data, make decisions, detect emergency and make prediction, perform data compression and encryption and adapt data capture frequency. They are able to contact the rescue teams according to the type of emergency detected and to notify the nearest hospital of the arrival of a patient (Figure \ref{fig_usecase2_house_plan}).  
These smart gateways make early recognition of deterioration more reliable by bringing the decision-making core and critical notifications closer to the patient. Due to the constant incoming data in a continuous medical monitoring, the system may encounter problems such as latency in system response, data transmission and computations related to data analytics. QoS has to be controlled from end to end. As previously shown, our SCC has been designed for that aim. As the user moves, service assembly is adapted in a way to follow him according to the requested QoS. The processing time of the services chain from end to end is thus controlled. This system takes advantage of local gateway processing to present notification and feedback with the minimum latency. The assembly is dynamic whenever the patient is moving. We compute an estimation of the processing time of the services assembly on a Raspberry Pi.

Each sensor is an SCC, each IoT gateway includes one or several SCCs and external services (hospital, rescue team) one or several SCCs too. To simplify, we consider only one SCC in an IoT gateway and in the external services.

The layout is here a pyramidal decreasing layout (Figure \ref{fig_three_layers}). It's often the case in IoT because data are processed and analyzed at each level. The result comes thus from aggregated data from several sources. The number of services at each level so decreases. These assembling layouts lead to well-known organizations like Edge \cite{art5:jain_fog_2016,art5:ahmed_mobile_2017}, Dew \cite{art5:wang_definition_2016} or Fog computing\cite{art5:datta_fog_2015,art5:sarkar_assessment_2015} \cite{art5:skala_scalable_2015} with three layers. 
Note that all calculations were done on a single raspberry pi. Calculations could be distributed on each device including SCC, which could considerably improve performance. 

We build the first layer with 10 SCCs (wearable user sensors) of type tA, 9 SCCs for the IoT gateways (Second layer, type tB) and five SCCs for hospitals (third layer, type tC) and two SCCs for the rescue teams (third layer, type tD) (Figure \ref{fig_three_layers}). Sensors have to be connected to one gateway. The gateway sends notifications to one hospital and one rescue team. Threshold values, link QoS and SCC processing time are random. Each IoT gateway SCC is able to accept 10 sensors SCCs simultaneous requests. Results are: number of SCCs: 26,
used memory: 280 KB, and processing time: 40 ms. Our algorithm is efficient for this use case. Assembly processing time is about 40 ms. So assembly can be done in a continuous way.
 
\textit{Discussion:} Implementation of such systems in real IoT scenarios faces significant/some challenges and constraints.
We highlight some of them and give some relevant references. 

As objects communicate with each other using an open channel, i.e., Internet, so security and privacy along with integrity of messages always remain a concern. 
Trust management among these objects is thus an open issue.
\cite{chahal_trust_2020} provide an elaborated view of trust management with a focus on Social IoT (SIoT) by comparing different existing trust management schemes based on the trust management process, parameters chosen for trust evaluation, characteristics of trust functions and objectives achieved by them.
  
Cognitive Internet of Things (CIoT) is another paradigm of IoT developed to enhance the capabilities of intelligence in IoT objects where these objects can take independent decisions in any environment.  
Unsolicited or undesirable electronic messages could disturb the internal cohesion of the assembly.
The power of learning, thinking, and understanding by these objects, can make the information access more accurate and reliable but Web spam is one of the challenges while accessing information from the web.
\cite{makkar_cognitive_2019} present an intelligent cognitive spammer framework which eliminates the spam pages during the web page rank score calculation by search engines.

In medical scenarios, due to the potential remoteness of the services, the latency (link QoS) between them has to be controlled (what our solution does) but has to be minimal too. 
The need of health care systems like telediagnosis, or robotic telesurgery is increasing day-by-day.
With these telesystems, doctors and patients need not to be present at the same location, but its requirements are fast response, ultra-reliable, and high availability. 
Traditional networks like 2G, 3G, and 4G-LTE are not able to fulfill these requirements, but \cite{gupta_tactile_2019} has shown that tactile internet and its applications in 5G era could solve this problem. 

\section{Conclusion}\label{sec_conclusion}

In this paper, we have proposed a self-assembling solution based on our self-controlled components that check the current behavior of the service and its conformity with the contract. 
This solution has several advantages.
The dynamic assembling core algorithm can potentially be run at anytime and can potentially be used as self-adjusting, self-adapting, and self-healing mechanisms.
For the user or architect, it simplifies fastidious manual assembling of a lot of devices. The service composition is made automatically without human intervention.
The architect can use the algorithm to simulate different organizations before deploying and implementing them in the real world
Our approach can build and maintain an assembly of services that, besides functional requirements, also fulfill global quality-of-service and structural requirements. Resource consumption and limitations of our approach have been reviewed. We have shown the feasibility of our approach on a case study. In a future work, we plan to deploy the proposed self-assembly algorithm on a full-scale real application scenario by taking into account the algorithm's energy consumption.

%

\section*{CRediT authorship contribution statement}
\textbf{Frédéric Lemoine:} Conceptualization, Methodology, Software, Validation, Investigation, Visualization, Writing - original draft, Writing - review \& editing. \textbf{Tatiana Aubonnet:} Supervision, Funding acquisition, Writing - review \& editing. \textbf{Noëmie Simoni:} Project administration, Supervision, Funding acquisition.

\section*{Acknowledgment}

This work is supported by the European Telecommunications Standards Institute (ETSI) project entitled: User-centric approach in the digital ecosystem (Specialist Task Force: STF BM/543).
The authors would like to thank Pr. Bernard Lemaire for his help in finalizing this paper.

\bibliography{biblio}

\end{document}